\documentclass[apj]{emulateapj}

\usepackage{natbib}
\usepackage{amsmath}
\usepackage{graphicx}
\usepackage{hyperref}
\usepackage{multirow}
\usepackage{threeparttable}
\hypersetup{backref,breaklinks,colorlinks,urlcolor=blue,linkcolor=blue,citecolor=black}

\newcommand{\Sersic}{S\'ersic}
\newcommand{\chisq}{\ensuremath{\chi^2}}
\newcommand{\chisqe}{\ensuremath{\chi^2_e}}

\newcommand{\RedeV}{\ensuremath{R_{e,\mathrm{deV}}}}
\newcommand{\IedeV}{\ensuremath{I_{e,\mathrm{deV}}}}
\newcommand{\PAdeV}{\ensuremath{\mathrm{PA}_{\mathrm{deV}}}}
\newcommand{\elldeV}{\ensuremath{e_{\mathrm{deV}}}}

\newcommand{\ReExp}{\ensuremath{R_{e,\mathrm{Exp}}}}

\newcommand{\FExpFtot}{\ensuremath{F_{\mathrm{Exp}}/F_{\mathrm{tot}}}}
\newcommand{\kms}{km s$^{-1}$}
\newcommand{\Mstar}{\ensuremath{M_\star}}

\graphicspath{{./}{figures/}}

\begin{document}

\title{Testing the presence of multiple photometric components \\
  in nearby Early-type Galaxies using SDSS
}

\author{
  Semyeong Oh\altaffilmark{1,3},
  Jenny E. Greene\altaffilmark{1} and
  Claire N. Lackner\altaffilmark{2}
}
\altaffiltext{1}{Department of Astrophysics, Princeton University, Princeton, NJ 08544, USA}
\altaffiltext{2}{Kavli Institute for the Physics and Mathematics of the Universe (WPI),
  Todai Institutes for Advanced Study, the University of Tokyo, Kashiwa, Japan}
\altaffiltext{3}{\tt{semyeong@astro.princeton.edu}}

\begin{abstract}

We investigate two-dimensional image decomposition of nearby,
morphologically selected early-type galaxies (ETGs). We are motivated by
recent observational evidence of significant size growth of quiescent
galaxies and theoretical development advocating a two-phase formation
scenario for ETGs.  
We find that a significant fraction of nearby ETGs show changes in isophotal shape
that require multi-component models.  The characteristic sizes of
the inner and outer component are $\sim 3$ and $\sim15$~kpc.  The inner
component lies on the mass-size relation of ETGs at $z \sim 0.25-0.75$,
while the outer component tends to be more elliptical and hints at a
stochastic buildup process. We find real physical differences between
the single- and double-component ETGs, with the double-component
galaxies being younger and more metal-rich.
The fraction of double component ETGs increases with
increasing $\sigma$ and decreases in denser environments. We hypothesize that 
double-component systems were able to accrete gas and small galaxies 
until later times, boosting their central densities, building up their outer 
parts, and lowering their typical central ages. In contrast, the oldest 
galaxies, perhaps due to residing in richer environments, have no remaining 
hints of their last accretion episode.

\end{abstract}    


\section{Introduction}

In the past ten years, starting with \citet{2005ApJ...626..680D},
there has been an increasing body of work showing that the sizes of
massive quiescent galaxies were, on average, significantly smaller
than their nearby descendants at fixed mass.  While the initial
discovery was based on rest-frame UV imaging, the rapid size evolution
of early-type galaxies (ETGs) has now been firmly established in the
rest-frame optical using near-infrared imaging
\citep{2006MNRAS.373L..36T,2007ApJ...656...66Z,2007ApJ...671..285T,
  2008ApJ...672..146S,2008ApJ...682..303M} and extended to large
\citep{2008ApJ...687L..61B} and spectroscopic
\citep{2008ApJ...677L...5V,2011ApJ...739L..44D,2011ApJ...736L...9V}
samples.  Recently, \citet{2014ApJ...788...28V} confirmed that the
mass-size relation of ETGs evolves rapidly over $0<z<3$ using a large,
spectroscopic sample from multiple fields.

Generally, this dramatic size evolution has been taken to mean that
individual massive quiescent galaxies have grown in size over the past
$\sim10$~Gyr. So far, there is no convincing observational evidence for 
a substantial population of compact quiescent galaxies
in the local universe at matching number density as the high-$z$ samples
\citep{2009ApJ...692L.118T,2010ApJ...720..723T}, although ongoing
searches for such galaxies both nearby and at intermediate redshift
may change this picture
\citep[e.g.,][]{2010ApJ...712..226V,2013ApJ...777..125P,
  2014ApJ...793...39D,2015A&A...578A.134S}.
The compact galaxies at high redshift must have
grown in size to transform into galaxies seen in the local
universe. Because the high-$z$ compact galaxies are often quiescent,
they are naturally imagined to be the progenitors
of present day ETGs, implying a significant size evolution
\citep[although see also][]{2015ApJ...804...32G}.

Alternatively, if increasingly large star-forming galaxies add to the
population of ETGs on the red sequence at later times, it is possible
that size evolution may appear more dramatic than it truly is for
individual systems
\citep[e.g.,][]{2013ApJ...775..106C,2013ApJ...773..112C,2012ApJ...746..162N}.
Disentangling the effect of this so-called progenitor bias from 
intrinsic size growth is a challenging task and is actively under
debate. If the progenitor bias is a more prominent cause of the size
evolution, on the one hand one expects to find some of the
undisturbed old, compact, and red galaxies in the nearby universe
as mentioned above. On the other hand, we also expect to find
consistently younger ages for larger ETGs which joined the population
at later times.  The recent study of \citet{2015ApJ...798...26K} looked 
at the stellar populations of $0.5 < z < 1.4$ ETGs as a function of 
their size and found interesting morphology-dependent trends 
\citep{2009ApJ...698.1590G,2010ApJ...712..226V}.

If there is indeed intrinsic size growth of compact red galaxies at
$z\lesssim 2$, what drives the evolution?  Several theoretical
scenarios have been proposed to explain the observed size evolution
including major \citep{2006MNRAS.369.1081B,2009ApJS..181..486H} and
minor mergers \citep{2009ApJ...699L.178N,2010MNRAS.401.1099H}, or
adiabatic expansion due to AGN feedback \citep{2008ApJ...689L.101F}.
One favored theory to explain such size evolution is a two-phase
formation of ETGs \citep{2009ApJ...699L.178N,2010ApJ...725.2312O}.  In
this scenario, a massive compact core forms predominantly in situ
through early star formation ($z>3$) and then in the second phase
grows by the dissipationless accretion of smaller galaxies over an
extended period of time, which accumulates mass in the outskirts and
increases the size rapidly ($R_e \propto \Mstar^\alpha$ where
$\alpha>1$).  A pure dry merger scenario, however, is challenged by
the lack of observed satellite galaxies to match the size evolution
\citep{2012ApJ...746..162N} and the non-evolving density slope of ETGs
measured from stellar kinematics and gravitational lensing
\citep{Sonnenfeld:2013aa}.  Recent cosmological simulations also find
that the evolutionary path taken by compact, massive
galaxies at $z\approx2$  can lead to a variety of outcomes
\citep{2016MNRAS.456.1030W}.  While about half of these systems
form the centers of
more massive galaxies today, some are completely disrupted, and others
remain compact, with some dependence on the environment.

In light of these recent developments, it is interesting to revisit
studies of the surface brightness profiles of local ETGs, although
traditionally they are often regarded as smooth and well-described by a
single photometric component.
If the present day ETGs are indeed formed in a two-phase
scenario, one may be able to see changes in the surface brightness 
profile where the outskirts were built up via minor mergers.

Several recent studies have performed multi-component fitting to
early-type galaxies in a similar context. \citet{Huang:2013aa} analyzed
optical images of 94 nearby early-type galaxies, and argued that the
majority of these galaxies in fact require three components to fully
describe their two-dimensional surface brightness profile.  They further
related the photometric substructure to a two-phase formation scenario
by comparing the mass-size relation and the surface brightness
profiles of the subcomponents with high-$z$ compact red galaxies
\citep{Huang:2013ab}, and argued that the decomposition reveals a
fossil record of mass accumulation on the outskirts of ``red nuggets''
at high redshift.  

\citet{2014MNRAS.443.1433D} also searched for faint
stellar halos around isolated central galaxies (not necessarily of
early-type) by stacking aligned galaxy images in bins of stellar mass
and concentration. The highest concentration, most massive galaxies in
their study (most relevant to our work) show a stellar halo reaching
to radii of $\sim 100$~kpc and a surface brightness below
30~mag~arcsec$^{-2}$. In their stacked images, the stellar halos
comprise a progressively larger fraction of the total flux, and
become more elliptical, as the stellar mass increases.  Finally,
\citet{2015MNRAS.446.3943M} performed multiple two-component fits
to the SDSS spectroscopic sample, yet focused on providing consistent
and accurate size or magnitude measurements across galaxies of
varying luminosity and morphological types rather than attaching
physical meaning to subcomponents, aside from noting that the classical
bulge+disk interpretation is probably not valid for the two component
galaxies at the brightest end.  These studies have not yet explored
any connection between the profile shapes and galaxy
stellar populations.

We note that the operating definition and selection of ETGs varies
between the studies discussed above. At high redshift, selections are
often based on galaxy color
\citep[e.g.,][]{2007ApJ...671..285T,2006MNRAS.373L..36T,2007ApJ...656...66Z,
  2014ApJ...788...28V} or Sersic index \citep{2012ApJ...753..167B}, while at low
redshift visual morphology, emission line strength, and/or concentration
could all be employed as well as color. The selection of ETGs based on
different galaxy properties may lead to different galaxy samples with
varying degrees of contamination from star-forming galaxies
\citep{2013A&A...558A..61M}. In fact, \citet{2015ApJ...798...26K}
found that subtleties in the sample selection of ETGs may lead to
different conclusions about the importance of progenitor bias.  When ETGs
are selected by color and bulge-to-total ratio only, massive compact
($R_e < 2$~kpc) ETGs are younger than their larger counterpart,
suggesting that the progenitor bias is not the main driver of the size
evolution.  However, when an additional morphological criterion of
smoothness \citep{2002ApJS..142....1S} is included, the outcome is
reversed.

Taking the idea that the compact, red galaxies at high redshift may
evolve to be nearby passive, quiescent ETGs, we
examine the photometric structure of morpholologically selected ETGs,
including morphological ellipticals with blue colors due to ongoing
star formation.  We thus decompose a nearby sample of
morphologically selected early-type galaxies into two photometric
components, envisioned to be a core that may have formed in an early
dissipational process and an outer component.  For the first time, we
also test for a relationship between stellar population ages and
multi-component structural fits.

The plan of the paper is as follows. We describe the sample and data in
\S\ref{sec:sampledata}, and justify our fitting method and model
selection in \S\ref{sec:fitting}. We present the results in
\S\ref{sec:results} and end with the summary and discussion in
\S\ref{sec:discussion}. Through out the paper, we adopt the concordance
cosmological parameters $H_0 = 70$ km s$^{-1}$ Mpc$^{-1}$, $\Omega_M = 0.3$,
$\Omega_\Lambda = 0.7$.

\section{Sample \& Data}
\label{sec:sampledata}

\begin{deluxetable}{lr}[htbp]
\tablecaption{Summary of Sample Selection\label{tab:sample}}
\tablehead{
  \colhead{Condition} & \colhead{Count}
}
\startdata
Parent sample \citep{2010ApJS..186..427N}             & 14034 \\
E/S0 without warnings                                 & 2497   \\
No pair/disturbed/tail/AGN flag with $\sigma>70$~km/s & 1365    \\
Valid physical parameters                             & 1134    \\
Good cutout images                                    & 877     \\
Further visual inspection \& cutout image requirement & 838
\enddata
\end{deluxetable}

\begin{figure*}[htbp]
  \centering
  \includegraphics[width=0.48\linewidth]{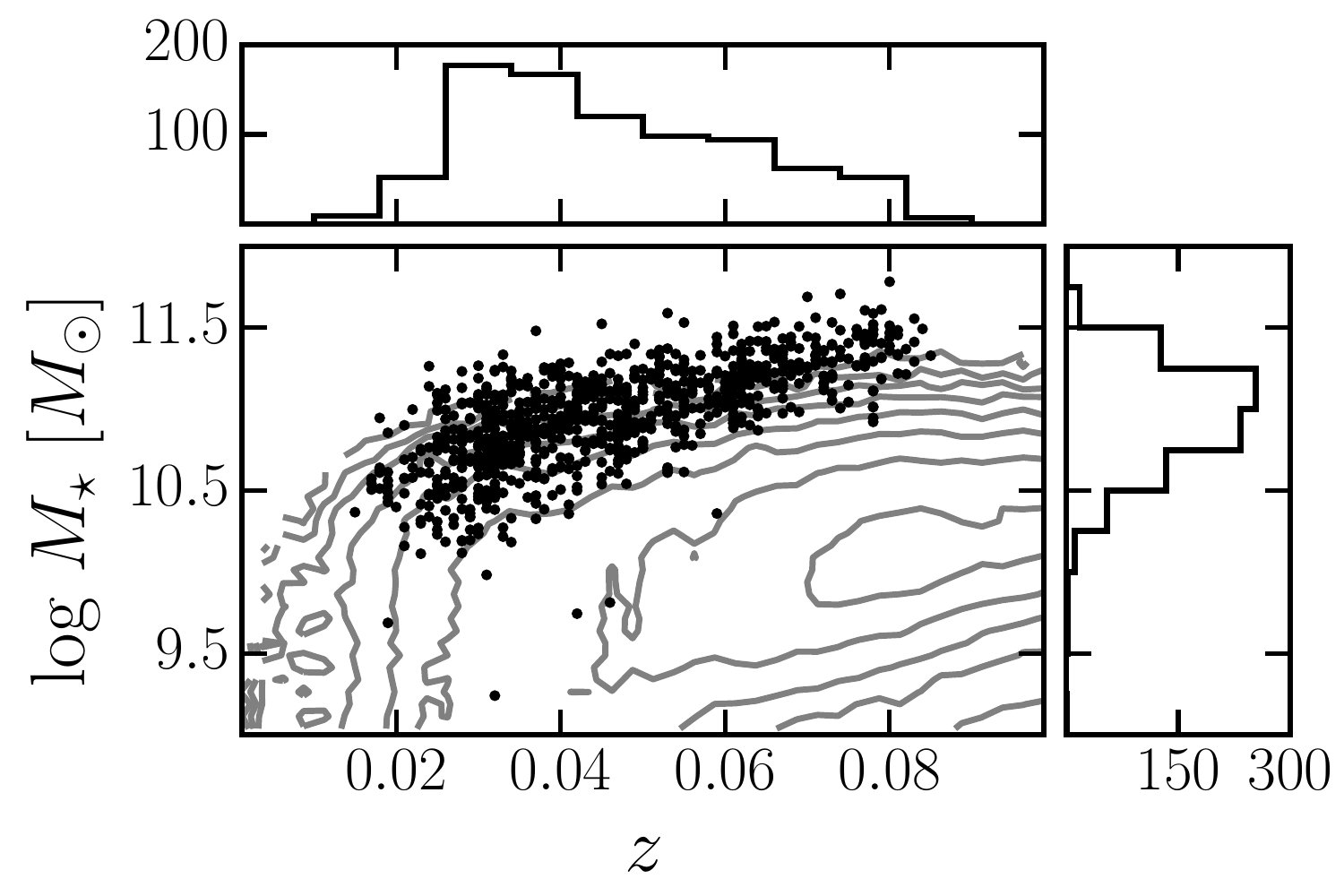}
  \includegraphics[width=0.48\linewidth]{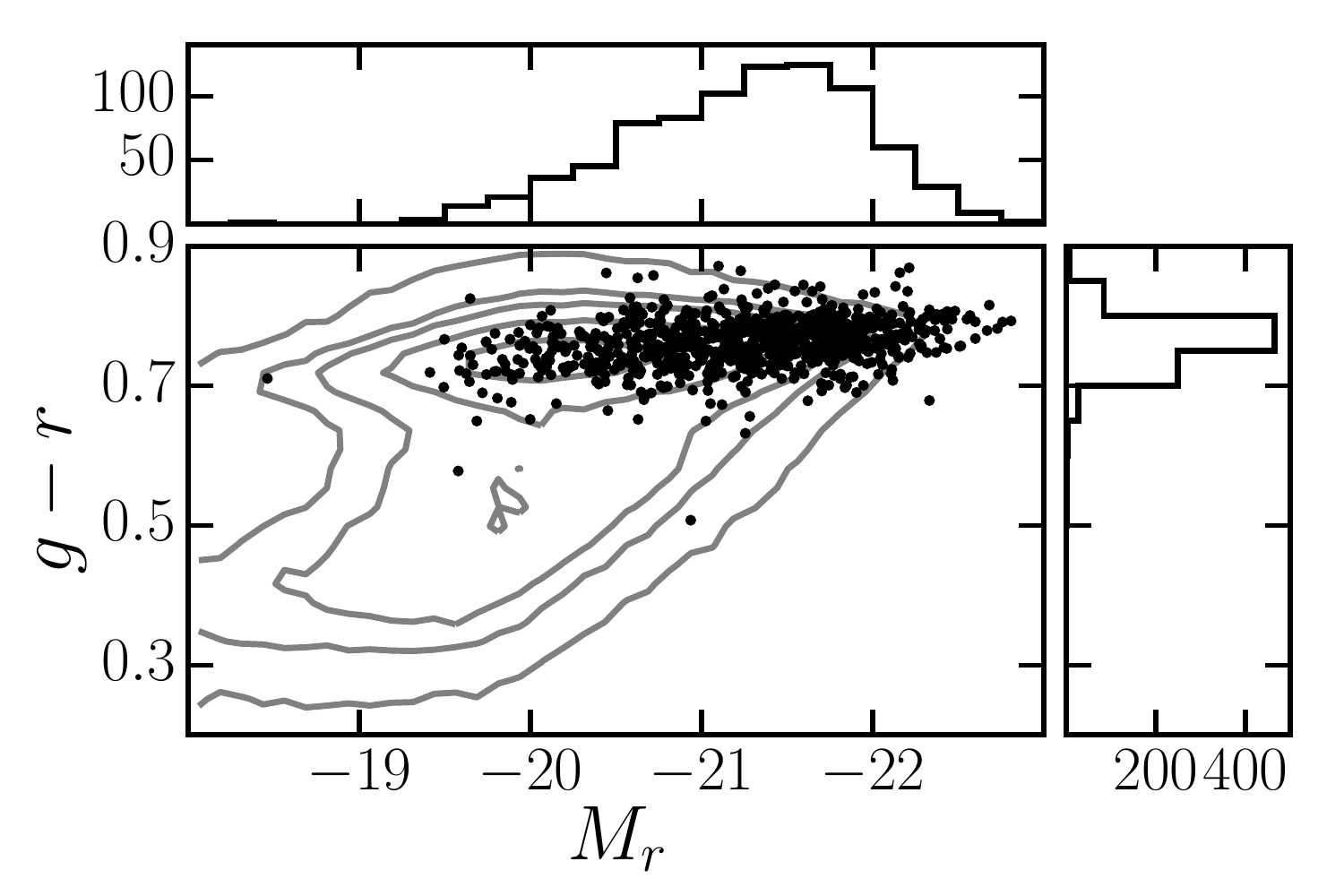}
  \caption{Distribution of the sample galaxies (black dots) in the redshift vs.
    $\log\Mstar$ plane (left) and $(g-r)$ vs. $M_r$ color-magnitude
    diagram (right). In each panel, we show the distribution of SDSS galaxies
    in the NYU-VAGC \citep{2005AJ....129.2562B}  at $z<0.1$ as gray contours
    with logarithmic spacing. Histograms show the distribution of each
    parameter for the sample only.
  }
  \label{fig:zmassrgr}
\end{figure*}

\begin{figure}[htbp]
  \centering
  \includegraphics[width=0.9\linewidth]{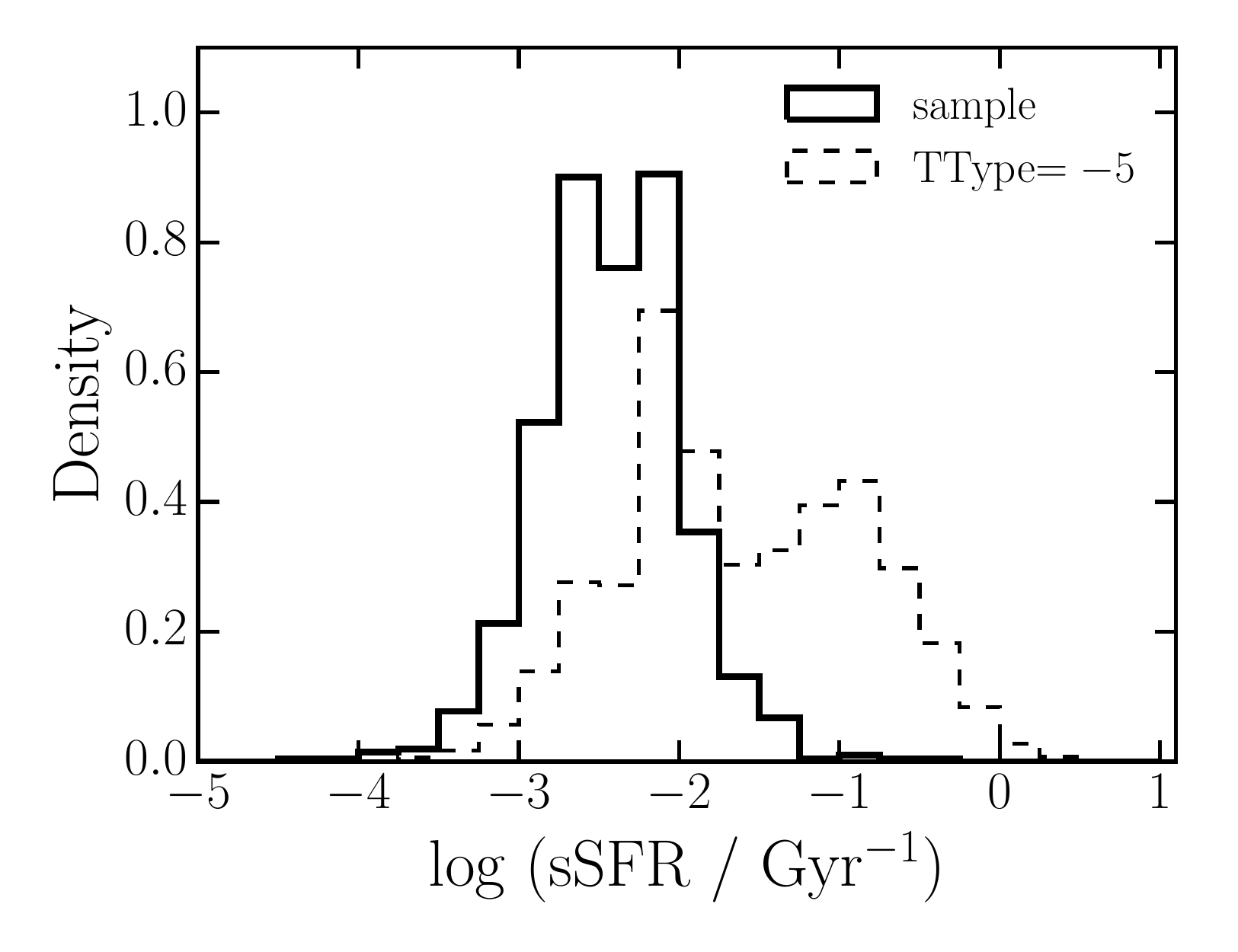}
  \caption{Density histogram of specific star formation rate of the sample. These
    values are derived from H$\alpha$ emission and color in
    \citet{2004MNRAS.351.1151B}. We show the distribution
    for all TType $=-5$ galaxies
    in \citet{2010ApJS..186..427N} for comparison. The galaxies in our sample have
    very low specific star formation rates, and does not include morphological
    ellipticals with specific star formation rate larger than $0.1$~Gyr$^{-1}$.
  }
  \label{fig:ssfr}
\end{figure}

We describe our sample selection and data used to construct cutout images
for fitting, which is summarized in Table~\ref{tab:sample}.
We use the detailed visual morphology classification catalog of
\citet{2010ApJS..186..427N} to construct a sample of nearby elliptical
galaxies in a range of mass and environment. The catalog is based on
the SDSS DR4 main spectroscopic sample, and extends to $z=0.1$ in 
order to include the most massive galaxies, which match most 
closely the mass of high redshift compact galaxies. We select unambiguously
identified RC3 Hubble types of cD, E, and E/S0
(TType $=-5$, flag on Ttype $=0$). Of the 2497 galaxies fitting these 
criteria, we remove galaxies flagged for disturbed morphology by selecting those
with disturbed flag $= 0$, and additionally inspect SDSS images visually to
remove galaxies showing any morphological peculiarities.
We also remove any galaxies with a close ``companion'' in the image (Pair =
0 and Pair flag type = 0) e.g., an overlapping star, for the same
reason.  We exclude galaxies harboring AGNs according to the classification
scheme of \citet{2001ApJ...556..121K} and \citet{2003MNRAS.346.1055K}
if this information is available.  We require that galaxies have valid
velocity dispersion measurements ($\sigma > 70$~\kms), stellar masses from
\citet{2003MNRAS.341...33K}, group halo masses from
\citet{2007ApJ...671..153Y} and an estimate of the environment
\citep{2006MNRAS.373..469B}.  We are left with
1134 elliptical galaxies that make these cuts. 

We use the SDSS $r$-band corrected frame images for surface brightness
profile fitting.  These images have been sky-subtracted by fitting a
smooth spline model, an improved method of estimating global sky
background compared to the previous PHOTO sky estimates
\citep[see][for details]{Blanton:2011aa}.  For the fitting, we cut out
frame images centered on each target galaxy with a size of 40
petrosian radii following \citet{2013MNRAS.433.1344M}.  We require
that more than 70\% of the desired cutout size be covered in a single
SDSS frame image.  We use {\it Source Extraction and Photometry in
  Python (SEP)}\footnote{{\tt https://github.com/kbarbary/sep}} to
make an object mask for each image. We mask all
neighboring objects of the target galaxy when fitting. As the central
part of the galaxy is important for surface brightness profile
fitting, we further require that no more than 10\% of the area in
concentric circles of radius 5, 10, 20, 50 pixels from the center be
masked.  Point spread function (PSF) images for PSF-convolution are
read from SDSS psField files at the position of each galaxy using the
SDSS {\tt readPSF} tool.
We visually check the integrity of
the remaining 877 galaxies with good cutout images
that satisfy the aforementioned requirements
using SDSS color composite images, and
exclude a small number of galaxies that are not elliptical
galaxies, or that suffer from artificial or astrophysical defects in
their images (Table~\ref{tab:sample}).

Our final sample with good cutout images comprises 838 elliptical
galaxies.  
Our sample spans a redshift
range of $0.015<z<0.085$ and a stellar mass range of $\log
(\Mstar/M_{\odot})=9.24$ to 11.78 (Figure~\ref{fig:zmassrgr}).
Although we do not specifically place a lower limit on stellar mass, we note
that 75\% of our sample galaxies have $\log\Mstar>10.75$~$M_{\odot}$. This
corresponds to the stellar-mass limit that significantly reduces contamination
in early-type samples from star-forming galaxies when using e.g., morphology to
select ETGs \citep{2013A&A...558A..61M}. Indeed, most of the galaxies in our
sample fall on the red sequence in the $(g-r)$ vs $M_r$ color magnitude diagram
(right panel of Figure~\ref{fig:zmassrgr}). We also investigate the specific
star formation rates (sSFR), as measured from H$\alpha$ emission and color
\citep{2004MNRAS.351.1151B}. The specific star-formation rates for our sample
galaxies are low (Figure~\ref{fig:ssfr}), with 86\% of the sample below
0.01~Gyr$^{-1}$. We note that of 5321 passive galaxies with
$\textrm{sSFR}<0.01~\textrm{Gyr}^{-1}$ in \citet{2010ApJS..186..427N}, 46\%
are ellipticals (TType = -5).
On the other hand, compared to all E/S0's visually classified by
\citet{2010ApJS..186..427N} (dashed line in Figure~\ref{fig:ssfr}), the
galaxies in our sample correspond to the most quiescent galaxies, and exclude
E/S0's with specific star formation rate higher than $0.1~\textrm{Gyr}^{-1}$.

\begin{figure*}[htpb]
  \centering
  \includegraphics[width=0.95\linewidth]{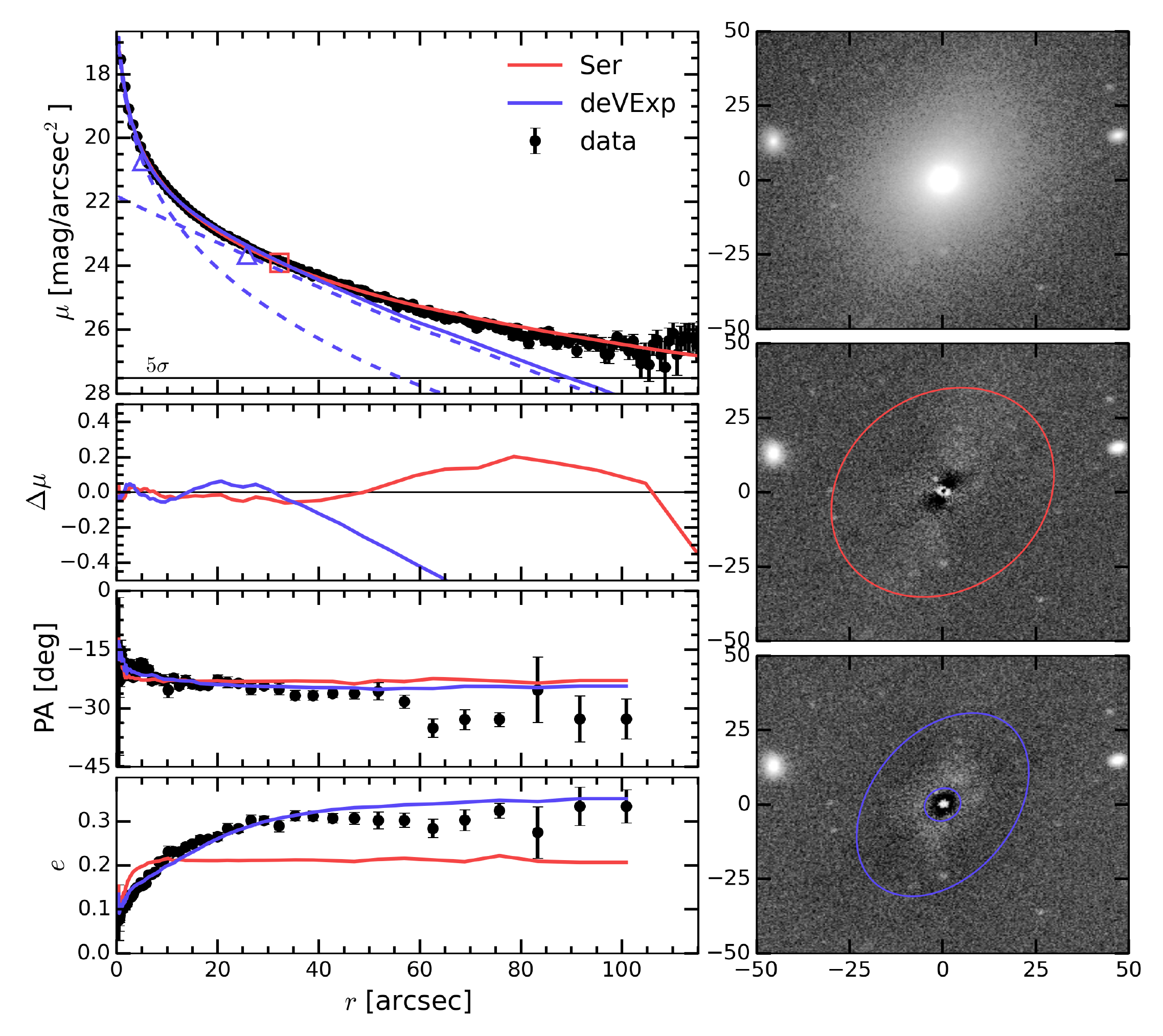}
  \caption{Example of a double component ETG selected with high confidence
    ($\Delta\chisqe>200$, sample id = 15). On the left, we show the azimuthal
    surface brightness profiles, residuals, PA, and ellipticity from top to
    bottom as a function of radius from the galaxy center for data (black
    circles) and two models, the single \Sersic\ (red) and deVExp (blue). On
    the top left panel, two blue dashed lines for deVExp model indicates the
    deV and Exp components respectively. For each subcomponent of the models,
    we indicate $R_e$ with markers of the same color. The horizontal line
    corresponds to the $5\sigma$ surface brightness limit. On the right, we
    show the galaxy image, and the residual images of the best-fit single
    \Sersic\ and deVExp model from top to bottom. The ellipses on the residual
    images indicate the effective ellipse for each component of the model.
  }
  \label{fig:exHC}
\end{figure*}

\begin{figure*}[htpb]
  \centering
  \includegraphics[width=0.95\linewidth]{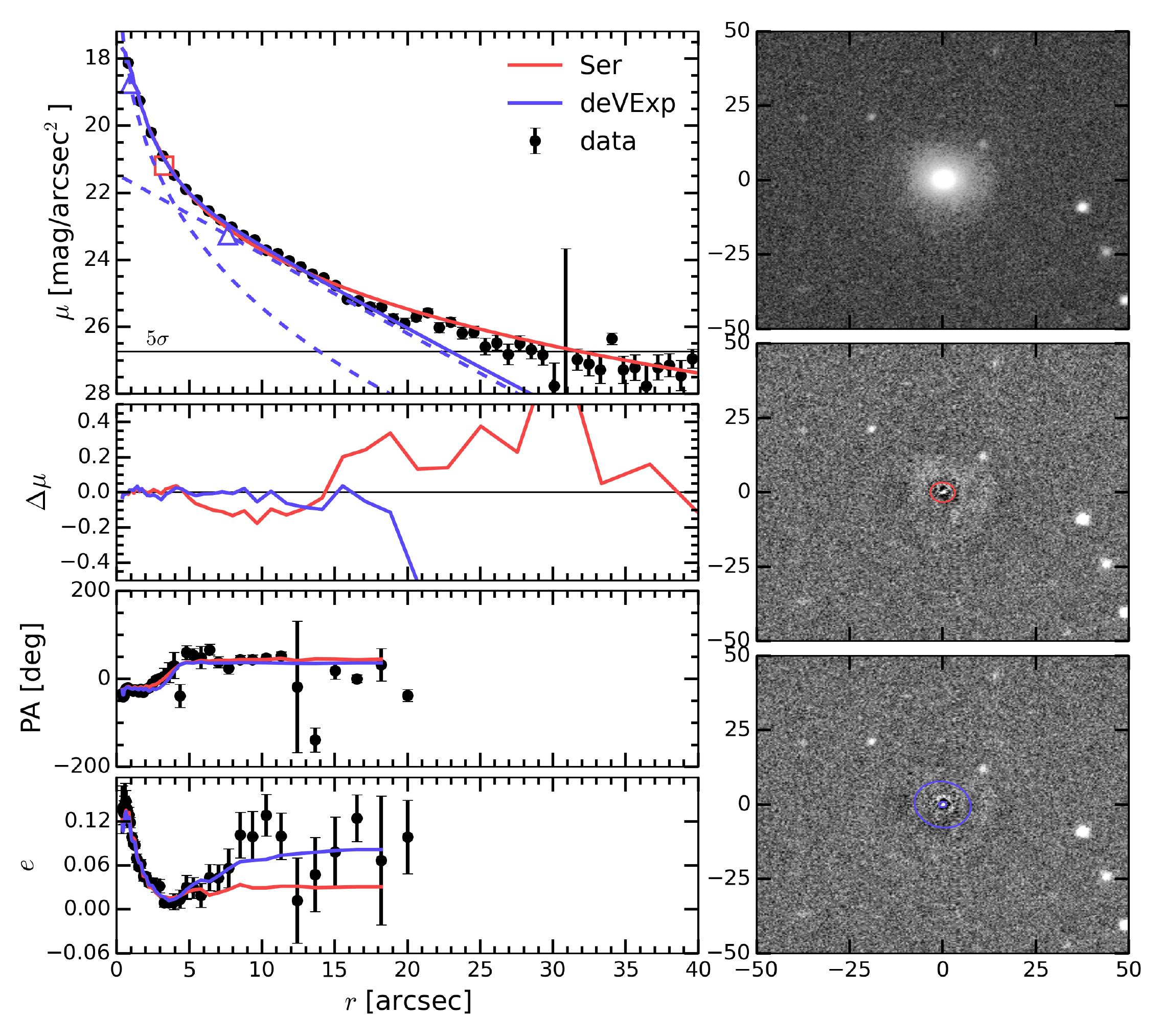}
  \caption{Same as Figure~\ref{fig:exHC} but for a double component ETG with
    $\Delta\chisqe<200$ (sample id = 188). At $r>20$~arcsec, the data are of
    insufficient signal-to-noise ratio to constrain PA and ellipticity.
  }
  \label{fig:exLC}
\end{figure*}

\section{Fitting}
\label{sec:fitting}

In this section, we describe our fitting procedure, including our choices of
model profiles and the fitting method (\S\ref{sec:procedure}), model selection
(\S\ref{sec:modelselection}) and the effects of PSF and residual sky background
to the fitting (\S\ref{sec:effects_of_sky_and_psf}).

\subsection{Fitting Procedure}
\label{sec:procedure}

We perform two dimensional image fitting using a combination of
\Sersic\ models \citep{1968adga.book.....S, Graham:2005aa}
\begin{equation}
  I(R) = I_e \exp \left(-b_n [ (R/R_e)^{1/n} - 1 ] \right)
  \label{eq:sersic}
\end{equation}
where the geometric shape of the isophotes are assumed to be ellipses
described by a center, position angle (PA), and ellipticity.
The ellipticity is defined as $1-b/a$ where $b/a$ is the minor-to-major
axis ratio of the ellipse.
Here, $R_e$ is the effective radius, $I_e$ the effective surface brightness,
and $n$ the \Sersic\ index.  The constant $b_n$ is chosen such that
half of the total flux is enclosed within $R_e$.  The allowed range of
the \Sersic\ index $n$ is between 0.1 and 9.  

We first perform the single \Sersic\ model fitting as a reference.
The two component model is a sum of a de Vaucouleurs and an
exponential with \Sersic\ indices of $n=4$ and $n=1$ respectively
(hereafter deVExp). The two components share a common center, but are
allowed to have different PA and ellipticity.  We choose to fix the
\Sersic\ indices of the two component model to minimize degeneracy
between index and size in \Sersic\ models \citep[e.g.,][]{2009ApJS..182..216K},
and place the sizes of each component among galaxies on equal footing.
Our goal is not to find the {\it best} model for the surface
brightness of each galaxy. Rather, we use the double component model
to test whether a two-component decomposition is justified.  We find
that this double component modeling is a convenient way to see if
changes in profile shape require this decomposition.  We are motivated
to use the exponential ($n=1$) profile for the outer part since it is
flat, making it less degenerate with the profile of the inner
component.  We further discuss our choice of indices and the effect of
freeing them in \S\ref{sub:correlations}.

We use IMFIT, a flexible galaxy image fitting tool
\citep{2015ApJ...799..226E}.  To find the best-fit model of a given
functional form for each galaxy, we minimize the residual sum of
squares using the Levenberg-Marquart gradient search algorithm.
Residuals are weighted by $1/\sigma_i$, where $\sigma_i$ is the error
at each pixel, obtained from the SDSS frame images.  For the deVExp
model, a choice of initial parameters may be important in reaching
optimal parameters.  While this algorithm only guarantees a local
minimum, we sample a grid of reasonable initial conditions to make
sure that we obtain the best model possible within the parameter space
of our interest.  

First, we fit a single de Vaucouleurs model. We use its parameters as
a reference for the deV component of the deVExp model.  The initial
parameters for the one-component fit are set using the existing
information on Petrosian half light radius and axis ratio from
\citet{2010ApJS..186..427N}.  While the true best-fit de Vaucouleurs
of the final deVExp model may change, the single-component fit still
serves as a good starting point.  Second, we fit the deVExp model with an
array of starting parameters.  Given the best-fit parameters \RedeV,
\IedeV, \PAdeV, \elldeV\ of the single deV model, we explore a grid of flux
ratios $\FExpFtot=0.1, 0.5, 0.9$ and size ratios $\ReExp/\RedeV=1, 5,
10$. In each of the nine combinations of these two parameters, we try
$\triangle \textrm{PA} = 0$ (parallel) and $90$ (orthogonal). We also try two
different initial ellipticities of 0.1 and 0.5 for the newly added Exp
component. This grid of values results in 36 different initial
conditions in flux ratio, size ratio, PA difference, and
ellipticity. After optimizing the model from this set of initial
conditions, we choose the one with the minimum best-fit chi-square
value as our best-fit deVExp model for each galaxy.

The azimuthal light profiles of nearby elliptical galaxies are
generally known to be well described by a single \Sersic\ model
\citep{1968adga.book.....S,1993MNRAS.265.1013C,2009ApJS..182..216K}.
Indeed, it is this flexibility that makes the \Sersic\ model excel in
fitting galaxy images, when accurately describing the photometric
substructure is not warranted (e.g., due to lack of resolution), or
not of interest. However, the definitive characteristic of a
two-component model that the single \Sersic\ fails to capture is any
change in isophotal shape reflected in e.g., PA or ellipticity as a
function of radius.  This has already been noted by e.g.,
\citet{Huang:2013aa} and \citet{2014MNRAS.443.1433D}, and was one of
their main reasons to prefer a multi-component model.

Adding the second (Exp) component, we find that the deVExp model adequately
describes the observed large scale shifts in isophotal shape. In many cases the
single-component fit fails to capture these changes. However, we often find
that the deVExp model systematically fails to account for the outermost light
profile. Usually the observed light profiles decline more slowly than an
exponential, typically at $\mu \lesssim 26$~mag arcsec$^{-2}$ and $r>60$
arcsec.

We show examples of ``double component ETGs'' that we determine are
better fit by the two-component model (deVExp), and we specify the
criteria below.  Figure~\ref{fig:exHC} is an example of a double
component ETG that we dub high confidence because the large shift in
ellipticity is much better captured in the deVExp model.  The
azimuthal surface brightness profile extends down to $26$~mag
arcsec$^{-2}$, and is well described by a single \Sersic. However,
this model fails to reproduce the ellipticity curve. The ellipticity
profile of the galaxy continues to rise to $e \sim 0.3$ at $5 < r <
30$, while the \Sersic\ model locks onto a value close to the best-fit
ellipticity. This is typical of single \Sersic\ model ellipticities
modulo some variation in the central region ($<5$~arcsec) due to the PSF. In
fact, one can see a hint of elongated low surface brightness light in
the residual image of the \Sersic\ model (middle panel of the right column
in Figure~\ref{fig:exHC}). The two component deVExp model, on the
other hand, can match the ellipticity profile by adjusting the outer
Exp component to have a higher ellipticity than the inner deV. As
discussed earlier, the decline in the surface brightness profile at
the very faint end is not as steep as exponential, and the deVExp
model fails at the largest radii. Considering only the 1D surface
brightness profile, one might think that the deVExp model could be
improved by increasing the $R_e$ of both components.  However, in 2D
fitting, the change in isophotal shape naturally demands the
arrangement of each component to simultaneously match the ellipticity
and PA profiles.

Figure~\ref{fig:exLC} shows an example double component ETG with
a modest change in ellipticity. The general characteristics are
similar.  Note that the non-trivial change in PA and ellipticity in
the inner $\lesssim 4$ arcsec due to PSF is quite well matched between
the data and the PSF convolved models.

\begin{figure}[htbp]
  \centering
  \includegraphics[width=0.9\linewidth]{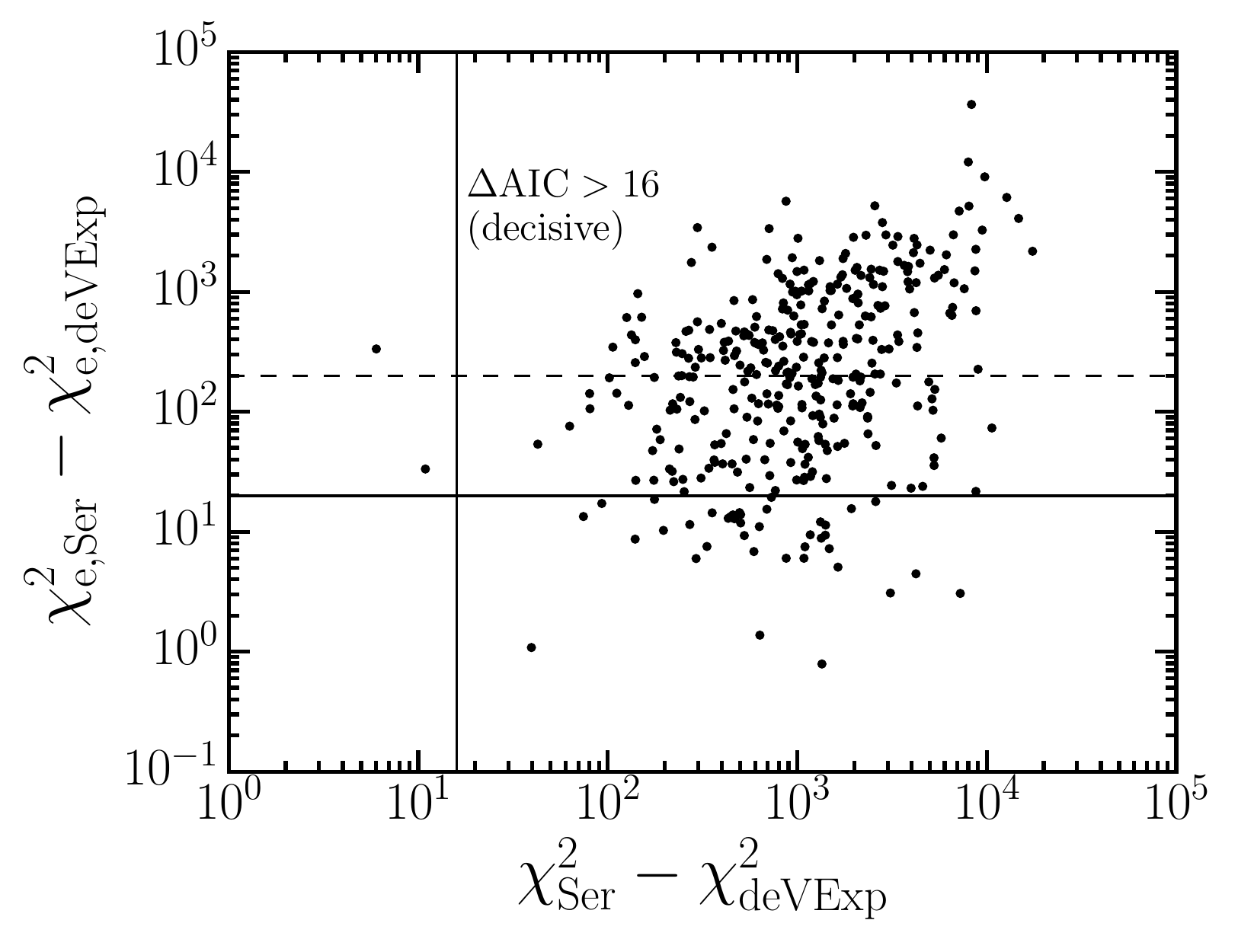}
  \caption{Correlation between the difference in  \chisq\ and \chisqe\ for ETGs
    not excluded by the conditions on Table~\ref{tab:flagnumbers}. The vertical
    line is a formal limit to prefer the deVExp model according to the
    Jeffreys' scale (See \S~\ref{sec:procedure} for a description). The solid
    (dashed) horizontal lines correspond to cuts for the assignment of double
    component (HC) ETGs. Note that the 65 (17\%) and 26 (12\%) of double and
    double HC ETGs for which the \chisq\ of deVExp model is increased relative
    to the \Sersic\ model are not shown.
  }
  \label{fig:ChisqDiff}
\end{figure}

\begin{figure}[htbp]
  \centering
  \includegraphics[width=0.9\linewidth]{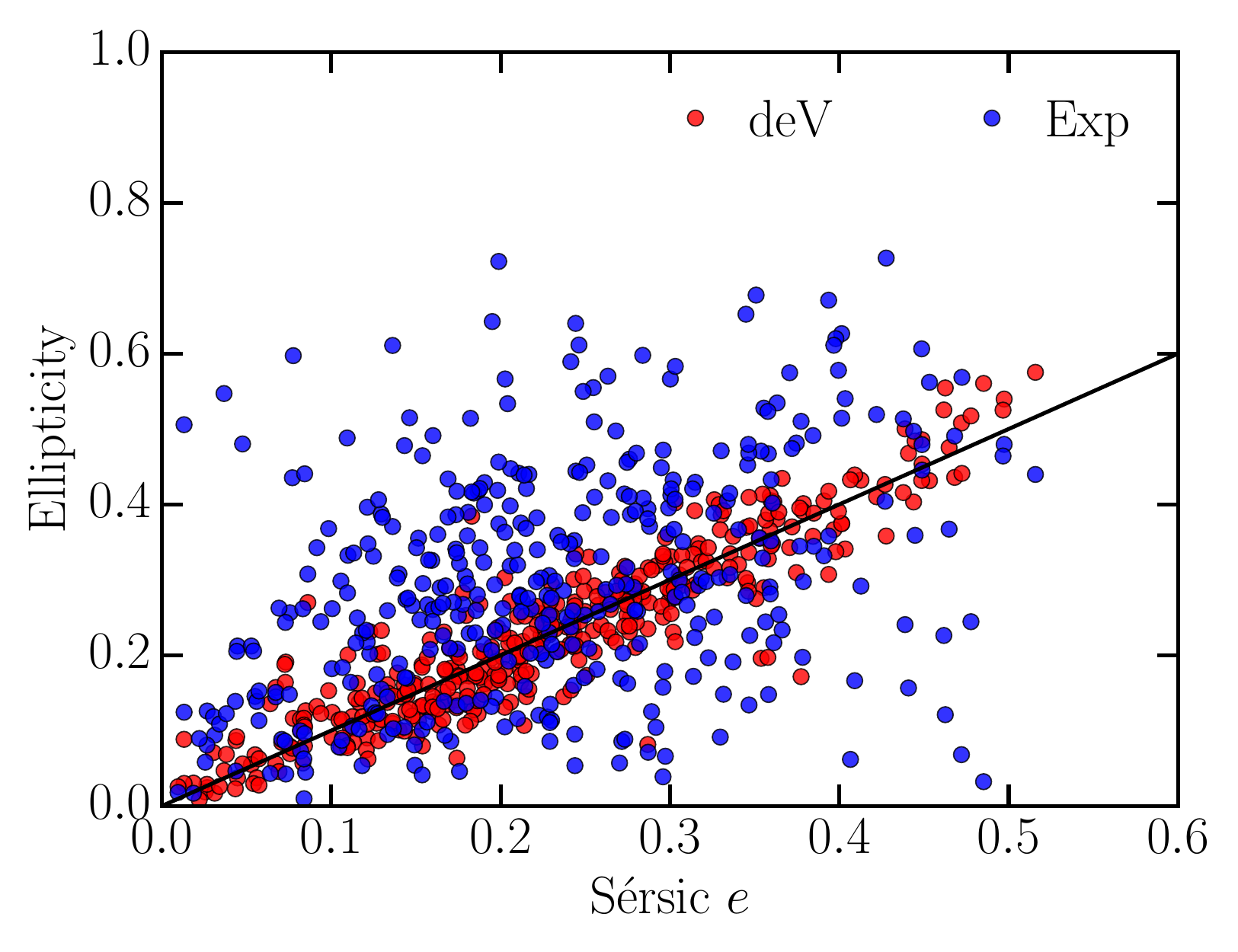}
  \caption{Component ellipticity vs. single \Sersic\ ellipticity for double
    component ETGs. Red and blue circles correspond to the deV and Exp
    ellipticity of the best-fit deVExp models.
  }
  \label{fig:ellipticity}
\end{figure}

\subsection{Model Selection}
\label{sec:modelselection}

With this in mind, we now turn to model selection between the single
\Sersic\ and the deVExp model.  Model selection is not an easy task in
general, and there is no absolute way to solve all model selection
problems.  Even in galaxy surface brightness fitting alone, we see a
variety of criteria adopted in recent studies:
\begin{itemize}
  \item \citet{2012MNRAS.421.2277L} use the $F$-distribution expected from the
    difference of \chisq\ to test the existence of
    a bulge component in addition to an exponential disk although the model is non-linear.
  \item 
    \citet{Huang:2013aa} selects models with the minimum number of components
    that are ``visibly distinct'', which primarily means that 
    not only the 1D surface
    brightness profiles but also the shapes (ellipticity and PA) 
    of isophotes should be recovered.
  \item \citet{2015MNRAS.446.3943M} uses an independent flagging system to
    place each galaxy into four categories, one of which is
    a two component galaxy. The final reliable two-component galaxies are
    those with ``intermediate bulge-to-total ratio and reasonably
    well-behaved subcomponents''. They, however, did not argue for a strong
    physical interpretation of the two component early-type galaxies.
  \item \citet{2014MNRAS.443.1433D} employed a Bayesian framework, and used a nested
    sampling algorithm to compute the Bayes factor when comparing single to multi
    \Sersic\ models for their stacked aligned images.
\end{itemize}
Perhaps the most conceptually straightforward and appealing approach
would be to do both parameter estimation and model selection with a Bayesian
framework.  However, this comes at the cost of considerable additional
computing power. We instead come up with a set of objective criteria
that are relatively simple to compute yet still take advantage of the
two dimensional fitting that we perform.  First, we discard the deVExp
model if the $R_e$ of the Exp component is smaller than that of the
deV, or is beyond the image size, in which case it is fitting the
small residual background.  Second, we require that the exponential
component dominates the surface brightness profile of the galaxy somewhere 
over the observed radial range.  

Finally, to select galaxies for which any change in isophotal shape is
better captured by the deVExp model, we compute the residual sum of
squares for each model from the ellipticity profiles.  We use the {\tt
  IRAF} ellipse task to fit the isophotes of galaxies with ellipses as
a function of semi-major axis for the data and both models.  Using
{\tt IRAF} ellipse, we are able to measure the PA and ellipticity down
to a mean surface brightness of 25.3~mag~arcsec$^{-2}$, at which
the signal-to-noise ratio is typically 25. Below this threshold, the
shape is unconstrained by the current data due to decreased
signal-to-noise.  We quantify the goodness of fit of each model
in describing the ellipticity curves as:
\begin{equation}
  \chisqe = \sum\limits_{a}\left(\frac{e_\text{data}(a) - 
      e_\text{model}(a)}{\sigma_{e,\ \text{data}}(a)}\right)^2
\end{equation}
where the summation is over the entire semi-major axis range, and
the radial binning of the data and the models are identical.
The meaningful quantity is the difference in \chisqe\ between the
single \Sersic\ and the deVExp model, 
$\Delta\chisqe = \chisq_{e,\mathrm{Ser}} -\chisq_{e, \mathrm{deVExp}}$.

After visual inspection of the residual images, the azimuthal surface
brightness profiles, and the PA and ellipticity profiles, we come up
with an empirical cut to select ``double component'' galaxies where
the deVExp model significantly improves the fit to any change in shape. We
prefer the deVExp model for any galaxy where $\Delta\chisqe > 20$, and
call this group double component ETGs.  We keep a subset of this group
with a higher $\Delta\chisqe$ threshold of $200$ ($\chisqe > 200$) as
high confidence (double HC) to assess the validity of our results on
double component ETGs with a more conservative subsample.
Table~\ref{tab:flagnumbers} summarizes our model selection process
with the number of galaxies flagged for each condition.

\begin{deluxetable}{lr}[htbp]
\tablecaption{Summary of Model Selection\label{tab:flagnumbers}}
\tablehead{
  \colhead{Condition} & \colhead{Count}
}
\startdata
Total                                    & 838   \\
a. \ReExp\ outside of image              & 45    \\
b. $\ReExp < \RedeV$                     & 94    \\
c. Exp subdominant everywhere            & 263   \\
Not excluded due to a, b, c              & 526   \\
$\Delta\chisqe > 20$ (double)            & 388   \\
$\Delta\chisqe > 200$ (double HC)        & 220   
\enddata
\end{deluxetable}

Although we select double component ETGs by \chisqe, we note that in
the majority of cases, the total \chisq\ is also significantly reduced
beyond the level that formally may be attributed to the increased
degree of model complexity. Figure~\ref{fig:ChisqDiff} shows the
distribution of the difference in the total chi-square,
$\Delta\chisq = \chisq_\mathrm{Ser} -\chisq_\mathrm{deVExp}$,
versus the
difference in \chisqe\ for those not excluded by the conditions on
Table~\ref{tab:flagnumbers}.  The strong correlation indicates that
the improvement in the double component model is accompanied by
improvement in fitting the change in ellipticity.  We use the Akaike
Information Criteria \citep[AIC;][]{1974ITAC...19..716A} to
demonstrate the improvement in \chisq. Under the assumption that the 
noise is normally distributed, the AIC is related to $\chisq$ through
$\mathrm{AIC} = 2 k + \chisq$ where $k$ is the number of free
parameters in each model (only differences in the AIC are meaningful,
not the absolute value).  According to the Jeffreys' scale
\citep{Jeffreys61}, one may take $\Delta\mathrm{AIC} > 10$ as strong
evidence to prefer the more complex model that reduces \chisq. Since
the deVExp model has three additional free parameters, the formal cut
to decisively prefer a deVExp model to the single \Sersic\ is 16,
which we show as a vertical line in Figure~\ref{fig:ChisqDiff}. We
note that 83\% (88\%) of our double component (HC) ETGs satisfy this
criterion, with most exceeding the cut by a large margin.

As already noted, often the azimuthal surface brightness profile of
ETGs falls off less steeply than exponential, which will tend to
decrease $\Delta\chisq$.  
Thus, only when there is a greater improvement in
fitting the overall surface brightness at 
$\mu\gtrsim 25$~mag~arcsec$^{-2}$, 
will the $\Delta\chisq$ will be positive by a large
margin.  Furthermore, it is worth noting that $\Delta\chisqe$
correlates with $\Delta\chisq$, hinting at a relation between
capturing the isophotal shape and the decrease in \chisq.
We also note that in 17\% (12\%) of double component (HC) ETGs,
\chisq\ is actually increased for the deVExp model, perhaps expected
from the occasional failure to fit the outer isophotes and the
flexibility of the \Sersic\ models.  We nevertheless include these in
our double component ETGs for consistency.

Figure~\ref{fig:ellipticity} compares the ellipticity of the deV and
Exp components with the ellipticity of the single \Sersic\ model for
each galaxy. It is clear that the inner deV components have almost
identical shape as the single \Sersic.  In fact, the difference in the
ellipticity of the inner deV and that of the single \Sersic\ is mostly
below $0.05$. This difference is expected since a single component
model will try to fit the shape of the central part of the galaxy
where the signal-to-noise is the highest.  On the other hand, the
outer Exp component is uncorrelated with the inner shape. This argues
that we are not just dividing a single \Sersic\ into two but actually
fitting outer light profiles of different shapes. It is also apparent
from this figure that the outer component tends to be more elongated
than the inner component.  We revisit the ellipticity distribution of
the subcomponents in \S\ref{sec:results}.

\begin{deluxetable*}{ll}
\tablewidth{0.8\textwidth}
\tablecaption{Structural parameters of
  bestfit models\label{tab:column_description}}
\tablehead{
  \colhead{Column name} &
  \colhead{Description}
  }

\startdata
ID & Unique sample ID \\
Name & SDSS name \\
RA & Right Ascension (J2000) \\
DEC & Declination (J2000) \\
$z$ & Redshift \\
$\sigma$ & Velocity dispersion in km s$^{-1}$ \\
$M_r$ & Absolute $r$-band magnitude \\
$g-r$ & Rest-frame $g-r$ color \\
FWHM$_\textrm{psf}$ & Full-width at half maximum of the PSF in arc seconds\\
$n_\textrm{Ser}$ & \Sersic\ index for single \Sersic\ model \\
$e_\textrm{Ser}$ & Ellipticity of single \Sersic\ model \\
$f_\textrm{Ser}$ & Flux of single \Sersic\ model in magnitudes\\
$R_{e,\textrm{Ser}}$ & Effective radius of single \Sersic\ model in arc seconds \\
flag & Flag on double component ETGs: 2 for high-confidence ($\Delta\chisqe > 200$) double component, \\
& 1 for double component with $20 < \Delta\chisqe < 200$, and 0 for single component ETGs  \\
$e_\textrm{deV}$ & Ellipticity of the deV component of deVExp model \\
PA$_\textrm{deV}$ & Position angle of the deV component of deVExp model in degrees \\
$R_{e,\textrm{deV}}$ & Effective radius of the deV component of deVExp model in arc seconds \\
$f_{\textrm{deV}}$ & Flux of the deV component of deVExp model in arc seconds \\
$e_\textrm{Exp}$ & Ellipticity of the Exp component of deVExp model \\
PA$_\textrm{Exp}$ & Position angle of the Exp component of deVExp model in degrees \\
$R_{e,\textrm{Exp}}$ & Effective radius of the Exp component of deVExp model in arc seconds \\
$f_{\textrm{Exp}}$ & Flux of the Exp component of deVExp model in arc seconds
\enddata

\tablecomments{
Table~\ref{tab:column_description} is published in its entirety in the electronic
edition of the Astrophysical Journal. A portion is shown here for guidance
regarding its form and content.
}
\end{deluxetable*}

We provide the derived structural parameters from the single
\Sersic\ and deVExp model (for double component ETGs) with other basic
properties of the sample in Table~\ref{tab:column_description}.

\begin{figure*}[htbp]
  \centering
  \includegraphics[width=0.9\linewidth]{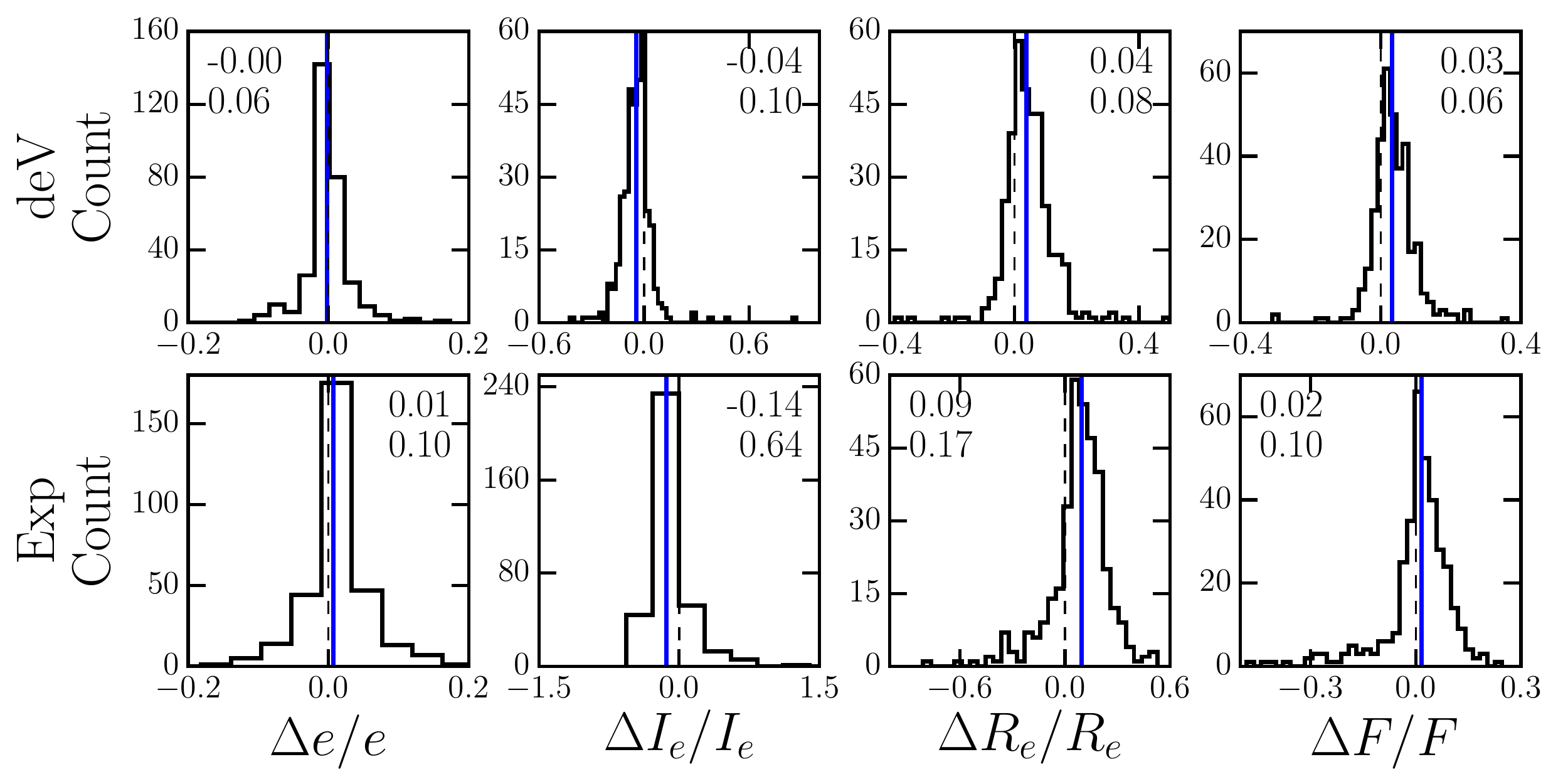}
  \caption{Histogram of fractional difference in key parameters of the best-fit
    deVExp model with and without a fixed residual sky component. The top and
    bottom rows show the distribution for the inner deV and the outer Exp
    component respectively. From left to right, the parameters are ellipticity,
    effective surface brightness, effective radius, and total flux. In each
    panel, the mean and the standard deviation of the distribution are quoted.
    The black dashed line and blue solid line indicate zero and the median
    respectively.
  }
  \label{fig:FracDiffSky}
\end{figure*}

\subsection{Systematic Uncertainty From Residual Sky and PSF}
\label{sec:effects_of_sky_and_psf}

The dominant potential sources of systematic uncertainty derive 
from errors in sky subtraction and the PSF model. 
We briefly discuss the effect of each to our
selection and parameter estimation of double component ETGs. 

As described in Section \ref{sec:sampledata}, we adopt sky-subtracted frames
from the improved algorithm of \citet{Blanton:2011aa}; here we only test the
impact of small errors in that sky value.
To test the effect of any residual
sky, we first estimate the residual by adding a flat sky component to the
single \Sersic\ model. We find that the magnitude of any residual sky
brightness in our cutout images is less than 25~mag arcsec$^{-2}$ and peaks
around 27.5~mag arcsec$^{-2}$.  We then add this value as a fixed flat sky
component in our modeling, and re-fit the deVExp model to see whether our
component determination changes.  We use the previously obtained best-fit
parameters as the initial guess.

\begin{figure*}[htbp]
  \centering
  \includegraphics[width=0.98\textwidth]{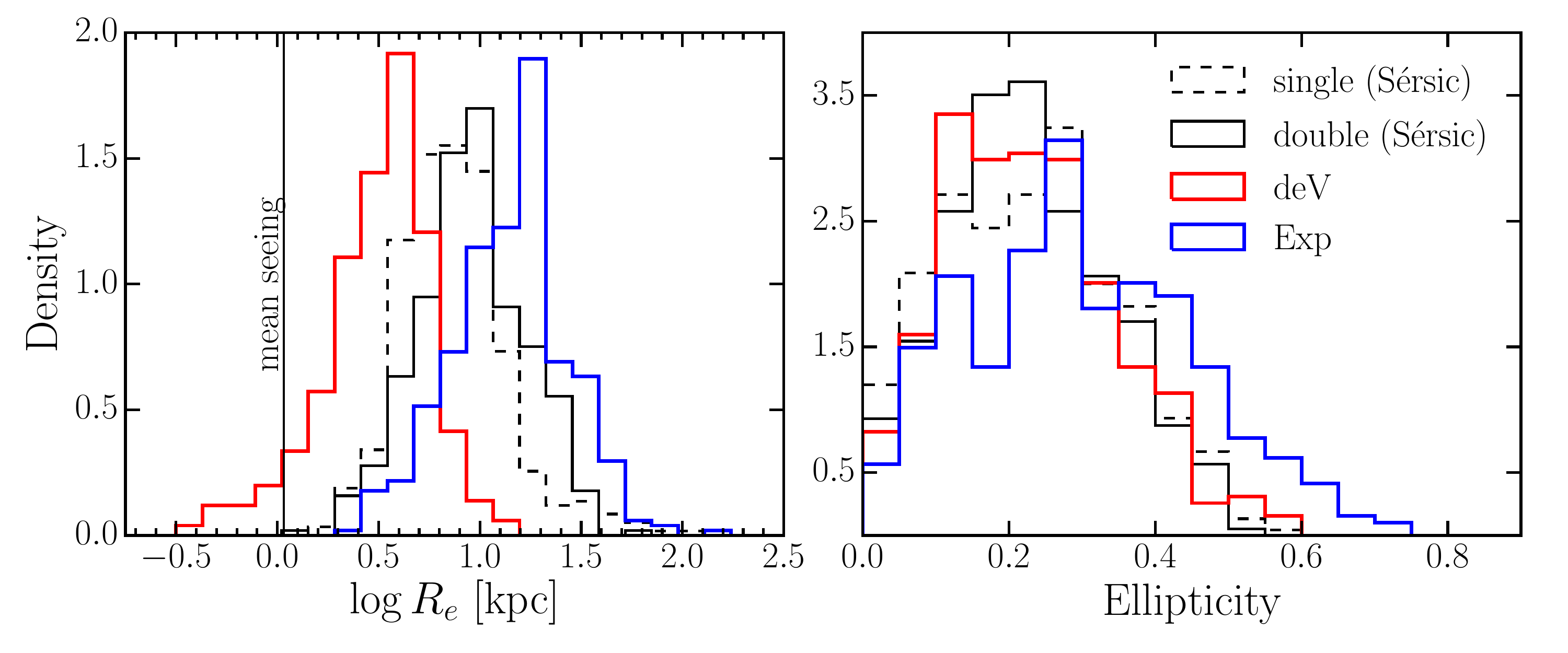}
  \caption{
    Distribution of size and ellipticity for single and double component ETGs.
    We present the effective radius and ellipticity measured using the single
    \Sersic\ model for single and double component ETGs in black, dashed, and
    solid lines respectively. For double component ETGs, we also show the
    distribution in $R_e$ and ellipticity measured for the deV (red) and Exp
    (blue) lines separately in the deVExp model. On the left panel, the
    vertical line marks the mean physical size corresponding to the seeing for
    the sample.
  }
  \label{fig:distSizeEllipticity}
\end{figure*}

Only a small fraction ($\lesssim 10$~\%) of our initial assignment to
single or double component ETGs change characterization when we change
the sky level. Of 388 (220) double component (HC) ETGs
(Table~\ref{tab:flagnumbers}), 35 (23) are excluded, and 25 (15) are
added when putting in a fixed flat sky component.  Given that this
fraction is so small, we conclude that residual sky errors do not
impact our conclusions. For simplicity, we maintain our
original classifications, but this test tells us that the systematic
errors in sky level do not impact our conclusions.

We also investigate how much the key fit parameters of each subcomponent change
due to sky errors using the subset of double-component galaxies with robust
classification. Figure~\ref{fig:FracDiffSky} shows the distribution of the
fractional differences in ellipticity, effective surface brightness, effective
radius, and total flux for the deV (top) and Exp (bottom) component.  We find
the position angle virtually unchanged, and omit this from the figure.  The
median of the distribution is marked with a blue line, and its numerical value
is indicated in each panel along with the standard deviation.

The ellipticities of the subcomponents are robust against
any residual sky, although the Exp shows the largest scatter.
Adding in a fixed sky has little effect on the inner deV profile,
resulting in an overall uncertainty of $\lesssim 10$~\% ($1\sigma$).
Unsurprisingly, the parameters of the outer Exp component show larger
deviations when a fixed sky component is included. The profile moves
systematically towards a slightly fainter but larger component. Although
the $I_e$ and $R_e$ of individual Exp components may vary by
$\approx20$\% or more, we find that the total flux in the Exp component is
much more robust (column 4 in Figure~\ref{fig:FracDiffSky}).  
The systematic errors due to sky presented here are dominant over
statistical errors. 
We estimate the errors in the parameters of either the single
\Sersic\ or deVExp models from the standard deviations of the
distribution of fractional difference in each parameter between the
model with no sky and the model with a fixed flat sky. For example, we
estimate the error on $R_e$ of the deV component in the deVExp
model to be 8\% (Figure~\ref{fig:FracDiffSky}).

Our selection of double component ETGs is unaffected by small
uncertainties in the PSF model.
The mean FWHM of the PSFs for the sample is $1.19 \pm 0.2$ arcsec. We note that
only 29 double component ETGs have an inner deV component that is only
marginally resolved ($ R_e < 1.5\times$ FWHM of the PSF).
As we show in \ref{sub:general_char}, 
in most cases, the inner deV component is well resolved.
The double-component ETGs that we probe in this work are those showing
changes in isophotal shape on kpc scales, beyond the central part of
the galaxy that may be most affected by any uncertainties in the PSF.

\begin{figure}[htbp]
  \centering
  \includegraphics[width=0.98\linewidth]{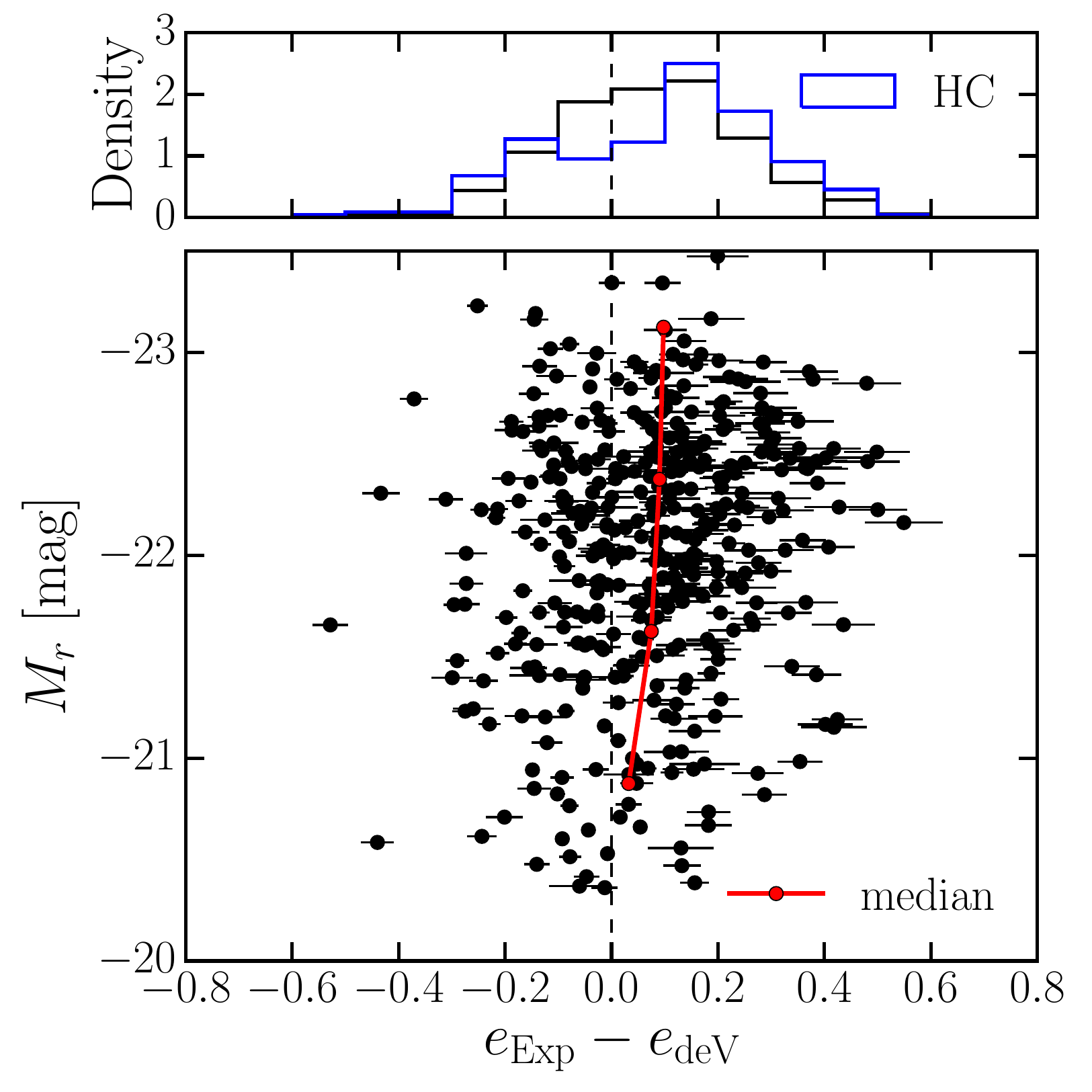}
  \caption{Difference in ellipticity of subcomponents vs. $r$-band absolute
    magnitude. The red circle connects the median in different magnitude bins.
    In the upper panel, we show the density histogram of the difference in
    ellipticities for double component ETGs, and highlight double component
    with high confidence with blue.
    }
  \label{fig:ediff}
\end{figure}

\section{Results}
\label{sec:results}

\subsection{General characteristics}
\label{sub:general_char}

Of the 838 elliptical galaxies in our sample, we find 388 double
component ETGs (46\%), 220 with high confidence (HC; 26\% of the
total). This number alone is interesting; a significant fraction of
nearby elliptical galaxies have complex photometric substructures that
cannot be well described by a single \Sersic\ model.

We first present the general characteristics of the subcomponents of
our double component ETGs.  For consistency, we measure all sizes in
the same manner. When referring to the size of a galaxy, we use the
single \Sersic\ fit, which in general provides the best overall fit to
the light profile. We will also consider the size (effective
semi-major axis) of each subcomponent for double component ETGs.

Figure~\ref{fig:distSizeEllipticity} shows the density histograms of the size
and ellipticity of the best-fit models for single and double component ETGs.
We first compare the sizes measured from a single \Sersic\ model for the single
and double component ETGs (black dashed and solid lines), and we do not find
any significant difference between the two groups.  For double component ETGs,
we also examine the component size and ellipticity for the deV (red) and Exp
(blue) components.  The inner deV and outer Exp component have characteristic
sizes of $2.3-4.6$~kpc and $9.2-20.5$~kpc, respectively ($25-75$ percentile).
Since our selection of double component ETGs is based on the radial change in
ellipticity, the change in isophotal shape is occurring between these two
characteristic radii.

We also compare the ellipticity of the two components in
Figure~\ref{fig:distSizeEllipticity}.  Since the shape of the deV is
essentially the same as that measured using the single \Sersic\ model
(Figure~\ref{fig:ellipticity}), we see no difference in the central
ellipticity for the single and double component ETGs.  In the double
component ETGs, the outer Exp components are found to extend to higher
ellipticity than the deV components. The median ellipticity 
for the deV and Exp components are $0.22$ and 
$0.29$, with standard deviation of $0.12$ and $0.15$, respectively.
Figure~\ref{fig:ediff} shows the distribution of the
difference in ellipticity between the deV and Exp component of the deVExp
model. On average, the outer component is more flattened than the
inner component with median $\Delta e \approx 0.082$. This difference
is more pronounced when considering only the double component HC ETGs
(median $\Delta e \approx 0.12$). The large scatter in ellipticity change
argues for a stochastic nature to the buildup of the outer component.

Our results are both qualitatively and quantitatively similar to the
findings of \citet{Huang:2013aa}, despite the differences in modeling
approach.  The outer component in their three component models, with a
median effective radius of $\approx 10$~kpc, is similar in scale to
the Exp component of our model. Their largest component also tends to
show higher ellipticities compared to the inner and middle at the
level of $\Delta e\lesssim 0.1$. They additionally find that the
skewness towards more elliptical profiles is most pronounced in the
high luminosity subsample ($M_V \le -21.3$ mag). We note that there is
a weak trend of increasing ellipticity difference with increasing
luminosity within our double component ETGs as well
(Figure~\ref{fig:ediff}; Pearson $r=-0.15$ with magnitude).
Considering that we intentionally selected ETGs that show significant
change in ellipticity, we also checked for a correlation between the
fraction of double component ETGs and the $r$-band magnitude, but do
not find such a correlation.

\begin{figure}[htbp]
  \centering
  \includegraphics[width=0.98\linewidth]{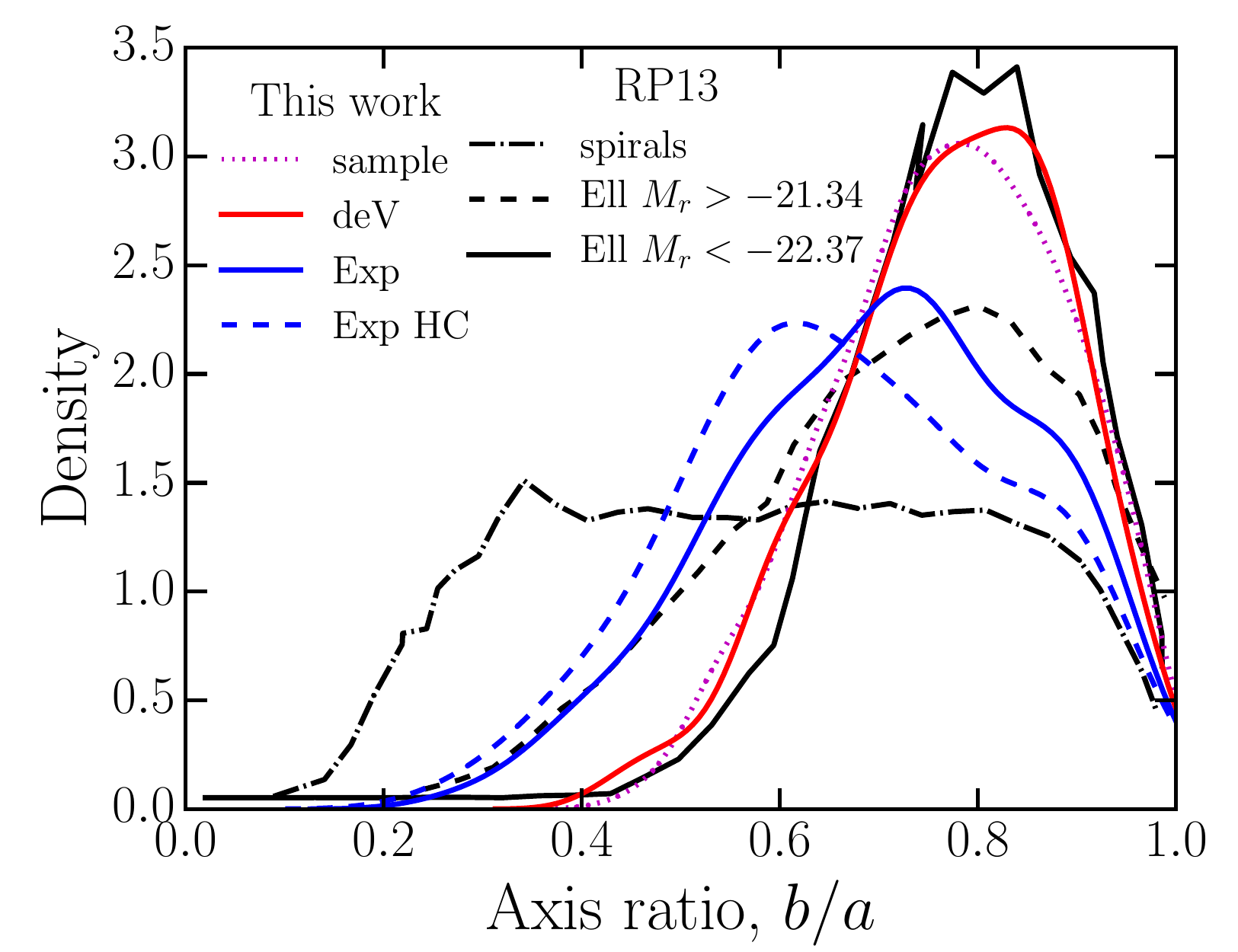}
  \caption{Axis ratio distribution of the sample and the double component ETGs.
    The axis ratio distribution for the entire sample from the single \Sersic\
    model (dotted magenta) and the axis ratios of each subcomponent, deV (red)
    and Exp (blue) of the double component ETGs are presented. We highlight the
    distribution for the Exp component of the double component ETGs with high
    confidence with blue dashed line. For comparison, we show the distribution
    for the highest and lowest luminosity bins of ellipticals, and for the
    spirals from \citet{2013MNRAS.434.2153R}. Density distributions are
    obtained from kernel density estimation using a Gaussian kernel.
    We find that the Exp component in the double component ETGs are more flattened than
    ellipticals, but their ellipticity distribution is still different
    from that of spiral galaxies,
    arguing that the outer components are not as flattened as exponential disks.
}
  \label{fig:axisratio}
\end{figure}

\begin{figure}[htbp]
  \centering
  \includegraphics[width=0.98\linewidth]{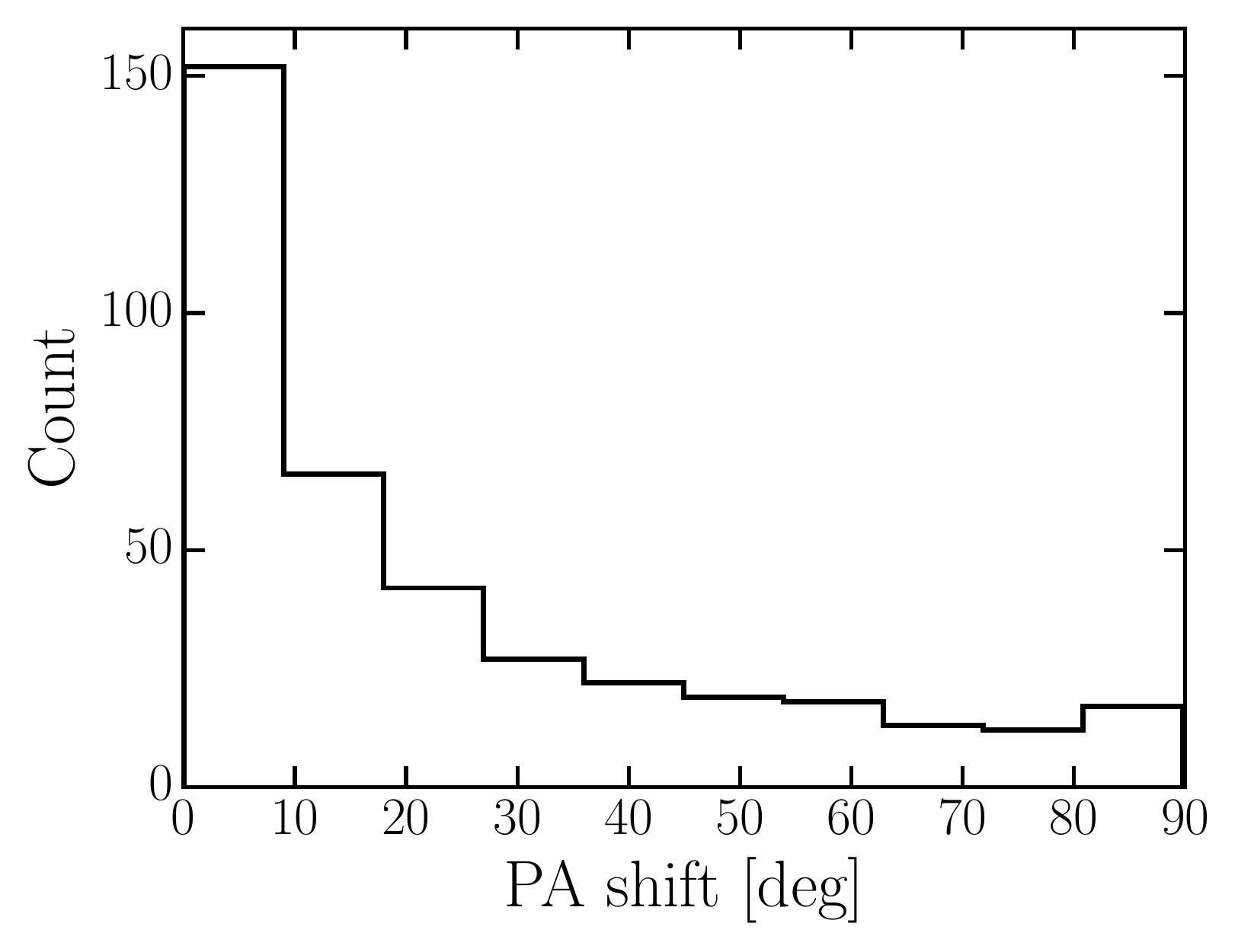}
  \caption{Histogram of the PA shift between the deV and Exp component of the
    deVExp model for double component ETGs.
    }
  \label{fig:pashift}
\end{figure}

The distribution in ellipticity for ensembles of stellar systems can
provide valuable clues to their intrinsic shape, which is ultimately
connected to the formation history \citep{1996AJ....111.2243T}.  We
compare the axis ratio distribution of our double component ETGs with
the latest morphology-specific galaxy axis ratio distribution from
\citet{2013MNRAS.434.2153R}.  Their study models the axis ratio
distribution of SDSS galaxies as the random projection of a triaxial
spheroid characterized by ellipticity $\epsilon = 1-B/A$ and thickness
$\gamma = C/A$, where $A$, $B$, $C$ are the major, middle, and minor
axes.  Elliptical galaxies show a variation in the apparent
ellipticity distribution with increasing absolute magnitude in the
sense that more luminous galaxies tend to be rounder
\citep{2013MNRAS.434.2153R}.  The variation is driven mainly by the
thickness ($\gamma$) that ranges from $0.584$ for the least luminous to $0.655$
for the most luminous group, while the ellipticity is always quite small
($\approx 0.1$).  Figure~\ref{fig:axisratio} compares the axis ratio
distribution of each subcomponent of the double component ETGs with
the least and most luminous ellipticals from
\citet{2013MNRAS.434.2153R}.  First, we find the overall distribution
of axis ratios of the entire sample to be consistent with
\citet{2013MNRAS.434.2153R} given that more than half of the galaxies
in our sample have an $r$-band magnitude brighter than $-21.9$ mag
(em3 and em4 in their paper).  For double component ETGs only, we
compare the distribution of each subcomponent to these references.
While the distribution of the inner deV component is very similar to
that of the entire sample and the most luminous ellipticals, the Exp
component shows a broader distribution, implying more flattened
systems.  In fact, compared to the least luminous ellipticals from
\citet{2013MNRAS.434.2153R} with $\langle\gamma\rangle = 0.543$ and
$\langle\epsilon\rangle = 0.122$, there is an indication that the Exp
component probed in this study would be even more flattened.  However,
we note that the distribution is still distinctively different from
spiral galaxies (dash-dotted line), from which we infer that we are
not simply identifying S0s.

We briefly compare our results with the two component decomposition of
\citet{2015MNRAS.446.3943M}. Of 682 galaxies that are common in both studies,
we find that the number of two-component galaxies using the same model (deVExp)
of \citet{2015MNRAS.446.3943M} is 399 (59\%). This higher fraction is probably
due to the less restrictive criteria of \citet{2015MNRAS.446.3943M}, which does
not explicitly utilize ellipticity profiles. The membership of two component
galaxy also differs in detail: 249 out of 399 two component galaxies are also
double component ETGs in our study while 57 of the double component ETGs in
this study are not labeled two-component by \citet{2015MNRAS.446.3943M}. The
general characteristics of size and ellipticity of their deVExp model are in
good agreement with our results. In their fits, the deV and Exp components have
typical sizes of $2.4-5$~kpc and $7-17$~kpc respectively ($25-75$ percentiles),
comparable to our values. They also find that the ellipticity distribution of
Exp component is more flattened than deV component, yet is not as flattened as
rotational disks.

We address the possibility that the observed change in shape in the
ETGs may actually be an isophotal twist in a single triaxial
component. The distribution of PA shifts between the two components is
shown in Figure~\ref{fig:pashift}.  Only 25\% of the double component
ETGs show a PA shift larger than 38 degrees between the two
components, and almost half of the double component ETGs are aligned
within 10 degrees.  As it is unlikely that the line of sight is along
the axis of a triaxial body in such large fraction of random
projections, triaxiality, although it may be of some importance,
cannot fully account for ellipticity changes
\citep{2005ApJ...618..195G}.  Another telltale sign of triaxiality is
a misalignment between the photometric and kinematic major axes.
\citet{2011MNRAS.414.2923K} investigated the distribution of this
misalignment for ATLAS$^{\rm 3D}$ ETGs and found that most (90 \%) of
the ETGs aligned within $15$ degrees, consistent with axisymmetry
within one $R_e$.

\begin{figure}[htbp]
  \centering
  \includegraphics[width=0.98\linewidth]{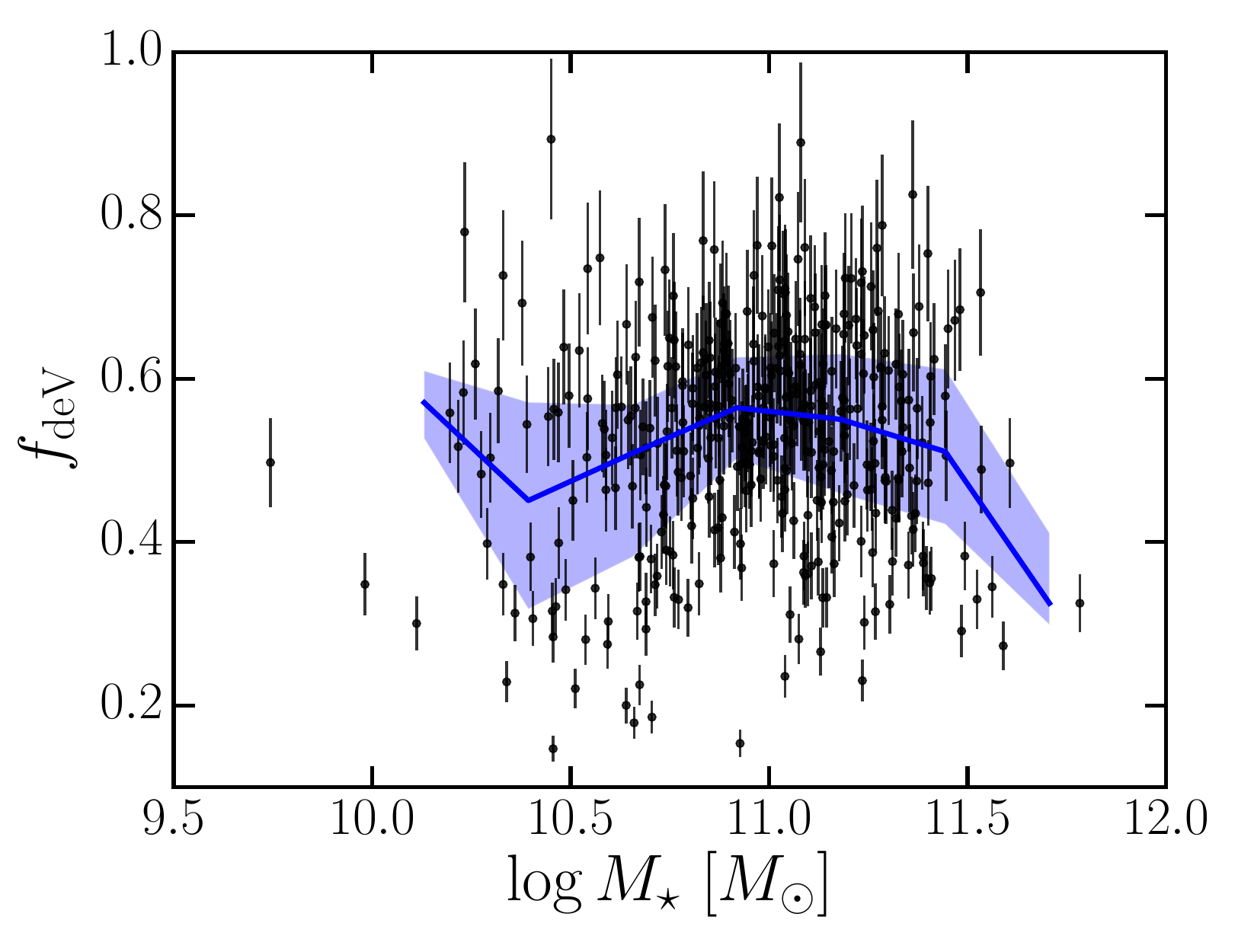}
  \caption{Flux fraction of inner deV component as a function of total stellar mass.
    We show the median and $25-75$ percentiles in equally spaced $\log\Mstar$ bins from
    10 to 12 with the blue solid line and shaded regions. 
  }
  \label{fig:fdev}
\end{figure}

We examine the flux fraction of the inner deV component in
Figure~\ref{fig:fdev}.  We use the single \Sersic\ model to estimate
the total flux, as the deVExp model misses some of the flux from the
outermost part (section~\ref{sec:fitting}). The fraction of flux in
the deV component $f_\mathrm{deV}$ shows a quite broad distribution
spanning $\sim 0.2$ to $0.8$ and typically is around $0.6$. This, in
turn, means that the fraction of light from the outer parts, which is
possibly material that is accreted after the core forms, shows a large
range. The flux fraction only shows a weakly increasing trend with the
total stellar mass (pearson $r=0.11$ for the whole sample).
A similar level of stochasticity is observed for
the outer components of the \citet{Huang:2013aa} decompositions (their
figure 29) for ETGs more massive than
$\log\Mstar>10^{10.5}~M_{\odot}$.

\begin{figure}[htbp]
  \centering
  \includegraphics[width=0.98\linewidth]{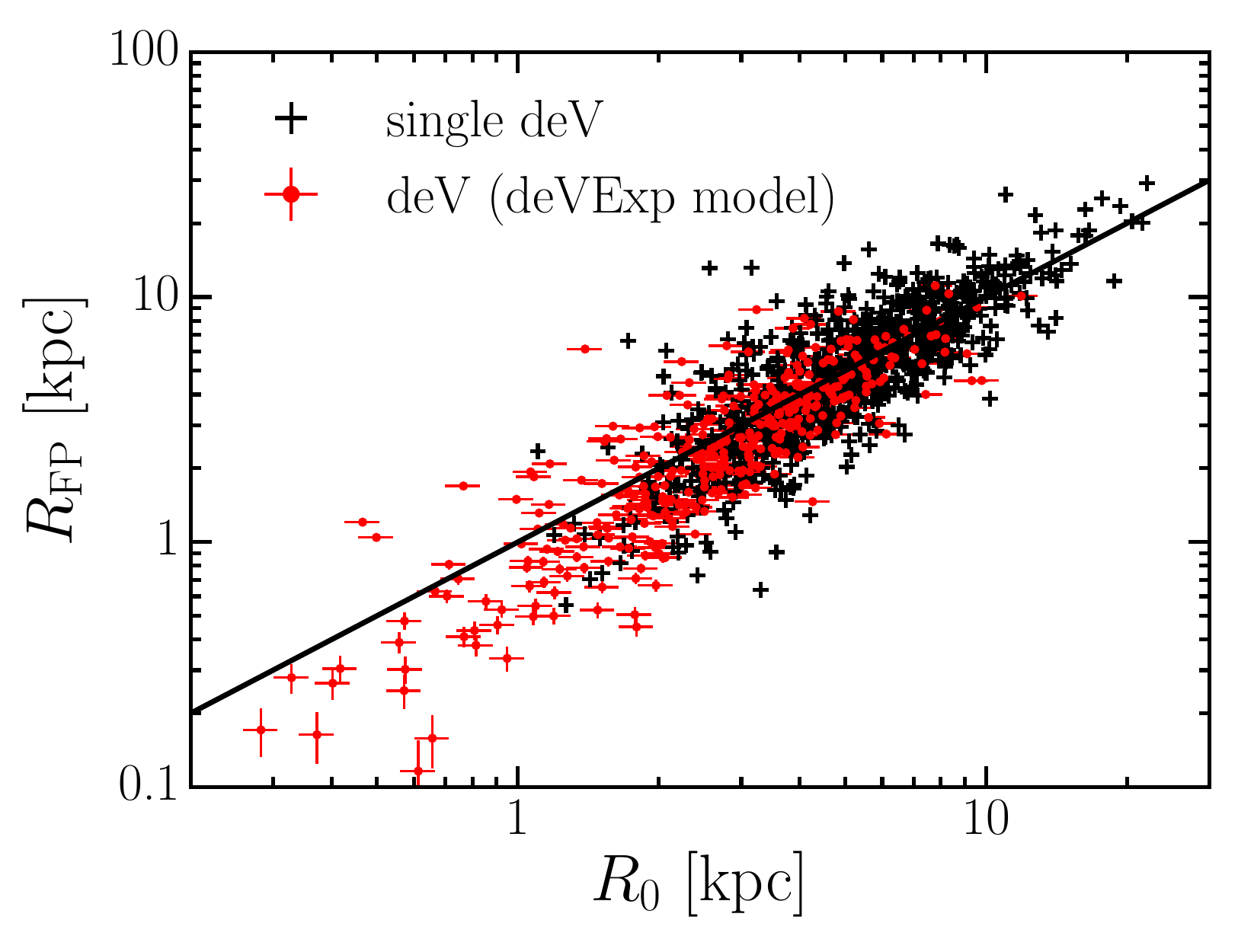}
  \caption{Distribution of the inner deV component of the best-fit deVExp
    models for double component ETGs on the fundamental plane (FP) defined by
    \citet{Saulder:2013aa}. We first check that the traditional single deV
    model parameters for our sample (black crosses) are consistent with their
    FP (black line). The inner deV component for double component ETGs (red
    circles) is shifted to smaller sizes, but moves along the FP.
  }
  \label{fig:fp}
\end{figure}

Elliptical galaxies show tight correlations between size, luminosity,
and $\sigma$, known as the fundamental plane
\citep[FP;][]{1987ApJ...313...42D,1987ApJ...313...59D,Bernardi:2003aa,Saulder:2013aa}.
We may ask whether the deV components of the double-component ETGs follow
the same scaling relations.  We take the FP coefficients from
\citet{Saulder:2013aa} for the SDSS $r$-band.  This relation was
derived for ETGs using the circularized radius $R_0 = R_e \sqrt{1-e}$
and the effective surface brightness $I_e$ from a single de
Vaucouleurs model fit. We first check for consistency using the same
model for the entire sample in Figure~\ref{fig:fp} (black points), and
find that the agreement is good.  For double component ETGs, we find
that the inner deV component essentially moves parallel to the
original FP with similar scatter.
The apparent deviation from the FP at small sizes ($R_0<1$~kpc) for the inner
deV components mostly correspond to marginally resolved cases with sizes
comparable to the PSF (see section~\ref{sec:effects_of_sky_and_psf}).

\begin{figure*}[htbp]
  \centering
  \includegraphics[width=0.98\textwidth]{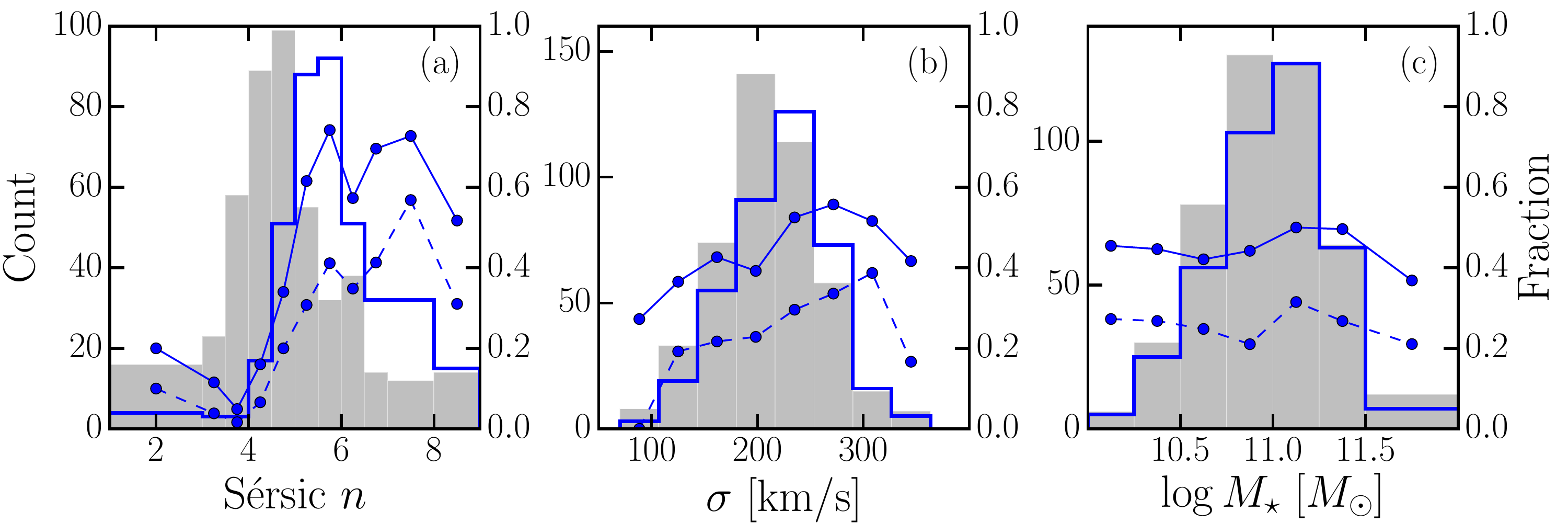}
  \caption{Histogram of galaxy parameters for single (gray bars) and double
    (blue steps) component ETGs. Solid (dashed) lines show the fraction of
    double component (HC) ETGs in each bin, with values on the right y-axis.
    The parameters are (a) \Sersic\ index $n$ from the best-fit single \Sersic\
    model, (b) galaxy velocity dispersion, and (c) photometric stellar mass.
  }
  \label{fig:HistGalaxyParams}
\end{figure*}

\begin{figure}[htbp]
  \centering
  \includegraphics[width=0.98\linewidth]{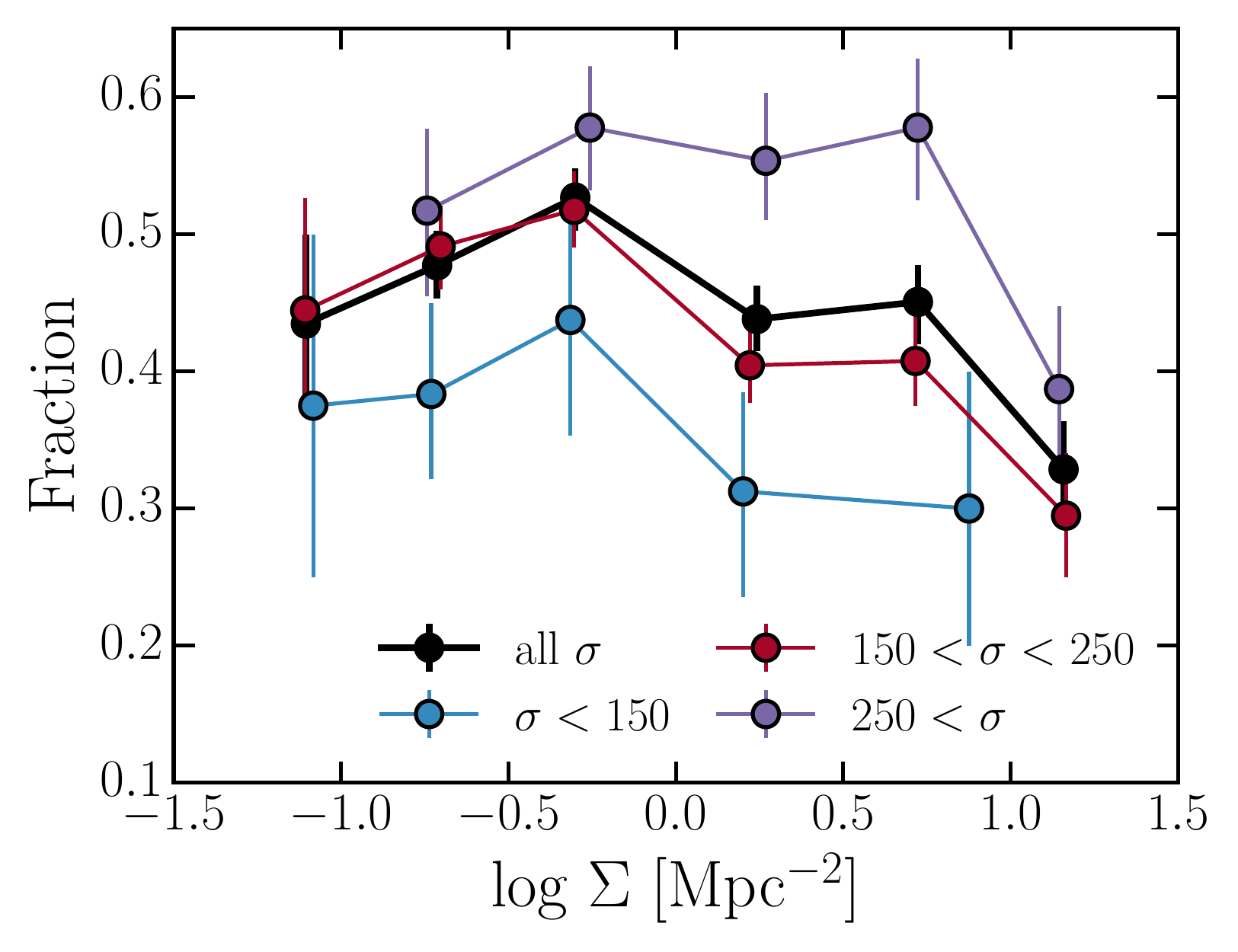}
  \caption{Fraction of double component ETGs as a function of projected
    neighbor density \citep{2006MNRAS.373..469B}. We show the trend in bins of
    $\sigma$. There is a weak trend towards a decreasing double component
    fraction with increasing galaxy densities.
  }
  \label{fig:density}
\end{figure}

\subsection{Correlations}
\label{sub:correlations}

To shed more light on the nature of the double component ETGs, we
explore correlations between the frequency of double
component ETGs and the intrinsic or environmental parameters of each
galaxy. Figure~\ref{fig:HistGalaxyParams} shows the
distribution of key intrinsic parameters for single and double
component ETGs. In each bin, we calculate the fraction of double
component (HC) ETGs, which are connected at the bin center by solid
(dashed) lines. By far the strongest trend we find is that high
\Sersic\ index ($n$) ETGs are more likely to be double component
(Figure~\ref{fig:HistGalaxyParams}a). Considering that the
\Sersic\ functions tend to be more centrally concentrated and more
extended at fixed surface brightness as $n$ increases, this trend may
seem trivial; a deVExp model will naturally work better when a galaxy
has a more extended outer surface brightness profile and thus high
$n$.  However, the important thing to note is that our selection
for double component ETGs is based on the ellipticity profile, or
isophotal shape. Thus, we find that ETGs with high \Sersic\ index 
are also more likely to show a radial change in ellipticity that is
better modeled by using a two component model.

It is possible that some of the low \Sersic\ $n$ galaxies also contain
a change in isophote shape, yet were mislabeled as single component
because we fix the inner index to $n=4$ and fail to find a good model
that simultaneously fits the overall radial profile and the shape.  As
a test, we fitted a double \Sersic\ model with both indices free
(between 0.1 and 9) to test whether our choice of indices were
reasonable, and whether our selection of double component ETGs changes
significantly.  We find that although there is a significant spread in
indices of both the inner and outer component,
they range from $n \sim 3$ to $\sim 6$ and $n <1$ to $\sim2$ for the inner
and outer component, respectively, not far from our choice of $n=4$
and $n=1$.  Since the double component ETGs in this work are selected
based on their ellipticity distribution, we do not expect our
selection of double component ETGs to be significantly altered. 
The component ellipticities of the double \Sersic\ model match our default
deVExp model with a scatter of $\sim 0.1$.
One should also keep in mind that the \Sersic\ $n$ depends sensitively on
how many \Sersic\ components are fitted.  For example, in the three
component model of \citet{Huang:2013aa}, most of the components have
$n\lesssim3$.

We also find an increasing fraction of double component ETGs with
increasing velocity dispersion [Panel (b) of
Figure~\ref{fig:HistGalaxyParams}]. This correlation is not
independent of the correlation with \Sersic\ index, as the velocity
dispersion correlates with \Sersic\ $n$ (Pearson $r=0.31$ for the
whole sample).  No similar trend is found with stellar mass [Panel (c)].
This may be related to the fact that velocity dispersion is a better tracer
of the central potential and the stellar population of ETGs in general than
the stellar mass
\citep{1993ApJ...411..153B,2000AJ....120..165T,Graves:2009aa,
  2012ApJ...751L..44W,2015ApJ...807...11G}.

Figure~\ref{fig:density} shows the fraction of double component ETGs
as a function of projected neighbor density from
\citet{2006MNRAS.373..469B}.  The projected neighbor density is
defined by averaging the surface density of galaxies within the
distance to the 4th or 5th nearest neighbor to a galaxy.  Since the
fraction correlates with the velocity dispersion $\sigma$, we divide
the sample into different $\sigma$ bins (colored lines), and also show
all $\sigma$ combined (black dashed line).  The error bars reflect the
25th and 75th percentile of the fractional distribution after
bootstrapping the sample 1000 times, assuming that our selection of
double component ETGs is correct.  We find that the fraction of double
component ETGs decreases significantly with increasing $\Sigma$ at
$\log\Sigma \gtrsim -0.5$, except for the highest $\sigma$ bin.

\begin{figure*}[htb]
  \centering
  \includegraphics[width=0.65\textwidth]{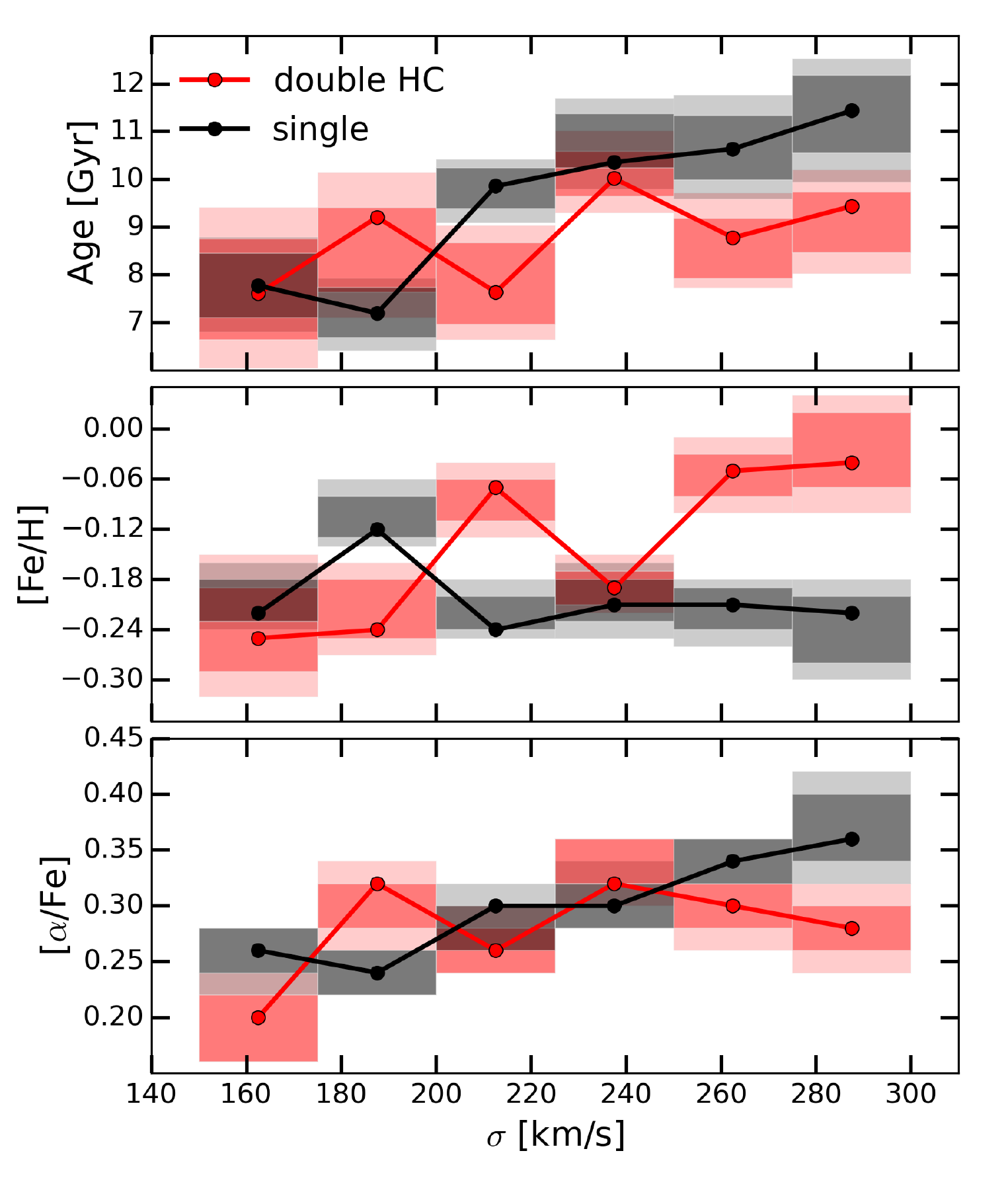}
  \caption{Stellar population difference between single and HC double component
    ETGs. Lines, dark and light shaded regions correspond to median, 70 and 90
    percentiles around median respectively from bootstrapping. The bins in
    $\sigma$ are chosen such that we have no fewer than 15 galaxies per bin. At
    $\sigma>200$~\kms\, we find that the double component (HC) galaxies tend to
    be younger and more metal-rich.
  }
  \label{fig:AgeAbundance}
\end{figure*}

\subsection{Stellar Populations}
\label{sub:stellar_population}

In this section, we use the SDSS spectra to look at the stellar
population properties of the single and HC double component galaxies.
Since it is well-known that the stellar population properties of ETGs are
a strong function of central stellar velocity dispersion
\citep[e.g.,][]{2000AJ....120..165T,Graves:2009aa,2014ApJ...791...45V}
we stack as a function of stellar velocity dispersion for both the
single-component and high-confidence double-component galaxies. We
also stack as a function of stellar mass. We have no fewer than 15
galaxies per stack. Note that, given the size
of the SDSS spectroscopic fibers (3 arcsec), we probe the central
stellar population such that light from the outer component of the
two-component ETGs does not contribute to the light in the fiber. 

Our procedure is as follows. First, we jointly fit the stellar
continuum and strong emission lines using pPXF+Gandalf
\citep{2004PASP..116..138C,2006MNRAS.366.1151S}.  We subtract emission
before stacking, but to be conservative we also remove the galaxies
with H$\alpha$ emission line EW$>0.5$~\AA, comprising $<10$\% of all
galaxies.

We then follow the stacking procedure described in detail in
\citet{Greene:2013aa,2015ApJ...807...11G}.  We divide each spectrum by
a heavily smoothed version of itself to remove the continuum and then
combine the spectra using the biweight estimator
\citep{1990AJ....100...32B}. We measure Lick indices from each stack
and use the code EZ\_Ages to invert the measured indices and
infer luminosity-weighted mean stellar ages, [Fe/H], and [$\alpha$/Fe]
\citep[for details see][]{2008ApJS..177..446G}. We derive errors
through a bootstrapping procedure in which we generate 200 mock stacks
drawing the same number of galaxies from the full list with
replacement.

The result is shown in Figure~\ref{fig:AgeAbundance}. We
compare the luminosity-weighted age, [Fe/H], and [$\alpha$/Fe] as a
function of $\sigma$ for single (black) and double (red) component
ETGs.  The median, 70\%, and 95\% of the distribution from the
bootstrapping is indicated with circles, darker, and lighter shaded
regions respectively.  While we find no difference at low $\sigma <
200$~km~s$^{-1}$, at higher dispersion the double component ETGs tend
to be younger and more metal-rich.  Although the difference in
$\alpha$-enhancement is insignificant in most bins, there is a hint at
the highest $\sigma$ bin that the double component ETGs may be less
$\alpha$-enhanced.  We find no such trends in stellar populations
when the spectra are binned
in stellar mass, although the range of stellar mass explored is narrow.

To test the importance of emission line subtraction,
we perform our analysis with no subtraction at all (clearly
unrealistic). While the ages move systematically by $\sim 2$~Gyr, the
difference in age between the two populations is preserved.

\begin{figure*}[htb]
  \centering
  \includegraphics[width=0.98\textwidth]{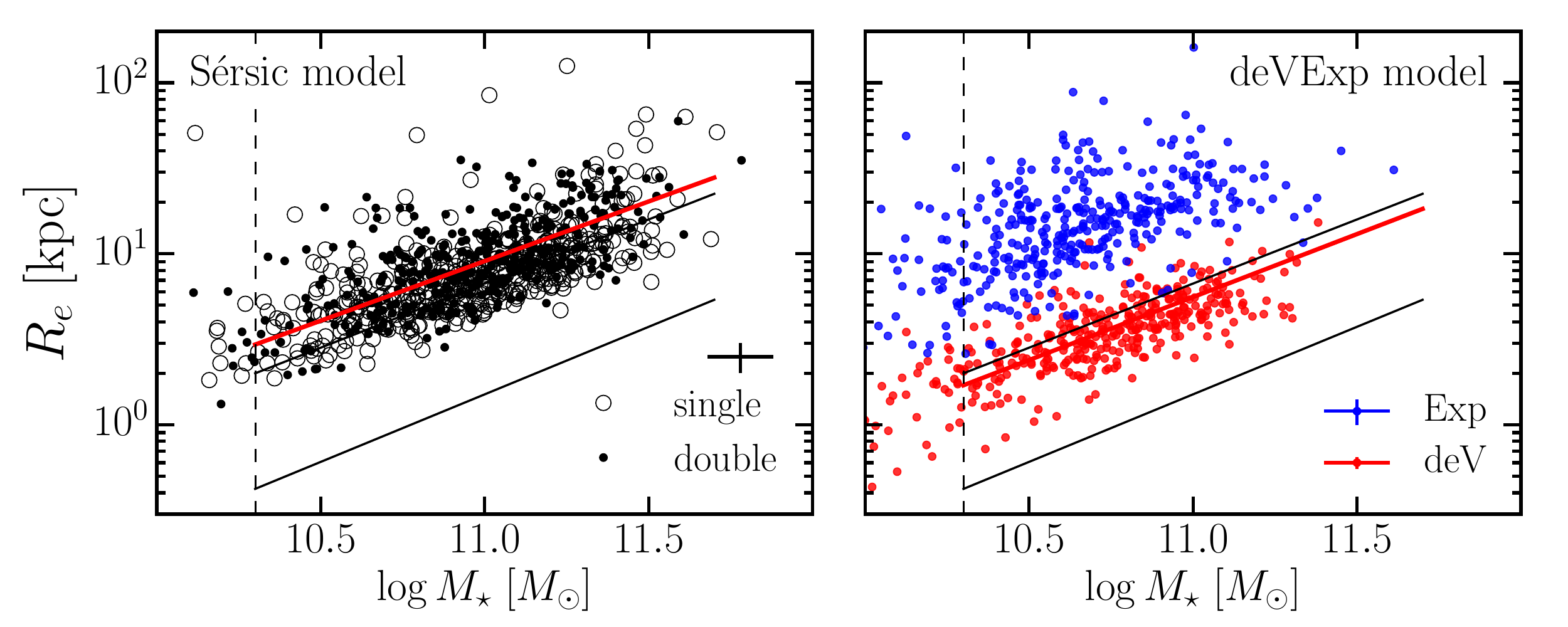}
  \caption{Mass-size relations of single and double component ETGs. On the
    left, we show the single \Sersic\ model sizes for the single and double
    component galaxies. We find that the double component ETGs do not separate
    from the single component ETGs on the mass-size plane. A power-law fit to
    the galaxies with $\Mstar>2\times10^{10}$~$M_\odot$ (vertical dashed line)
    is shown in the red line. On the right, we show the mass-size relation for
    each component of the deVExp model for double component ETGs. Similarly, we
    show a power-law fit to the deV component in red. In each panel, the upper
    and lower black lines show the mass-size relations from
    \citet{2014ApJ...788...28V} at the lowest and highest redshift bins
    ($z=0.25$ and $z=2.75$), which encloses the entire range from the present
    day to $z=3$. The typical errors in $x$ and $y$ are indicated in the lower
    left.
  }
  \label{fig:MassSize}
\end{figure*}

\subsection{Mass-Size Relations}
\label{sub:mass_size_relations}

In this section, we explore the mass-size relation of our sample
relative to literature that looks at the evolution of the
relation from $z=3$ to the present. For comparison, 
we take the observed mass-size
relations for quiescent samples from \citet{2014ApJ...788...28V} between 
their highest ($z=2.75$) and lowest ($z=0.25$) redshift bins 
to summarize the evolution in the mass-size relation from $z=3$ to the present.
Their sizes are measured by
modeling \emph{HST} images with a single \Sersic\ function
\citep{2012ApJS..203...24V}, and converted to size at the wavelength
of $5000$~\AA.

We first compare the equivalent single \Sersic\ model sizes from our
sample with the $z=0.25$ distribution, shown in the left panel of
Figure~\ref{fig:MassSize}. As discussed previously in \S
\ref{sub:general_char} and \ref{sub:correlations},
the single and double component galaxies have 
similar size and mass distributions, and the loci of the two
groups coincide in the mass-size plane. This also provides a
consistency check. As expected, the loci of our single-component 
fits lie slightly above
those of the $z=0.25$ galaxies (lookback time $\approx 3$~Gyr).
For comparison with van der Wel, we perform a power-law fit to the
single \Sersic\ sizes for galaxies with $\Mstar > 2\times
10^{10}$~$M_\odot$ \citep[consistent with ][]{2014ApJ...788...28V}
with a random error of 20\% in $R_e$ and 0.1 dex in
$\log\Mstar$ \citep{2007ApJ...657L...5K}.
We fit for $A$, and $\alpha$ in $R_e/\text{kpc} = A
(\Mstar/5\times10^{10}~M_\odot)^\alpha$.  The best-fit values of $\log A$ and
$\alpha$ are $0.73 \pm0.02$ and $0.70\pm0.05$.  We find that the
relation for our sample at the median redshift of 0.043 is consistent with
that of the quiescent galaxies at $z=0.25$ in
\citet{2014ApJ...788...28V} with a slight offset towards larger sizes.

For double component ETGs, we compare the mass-size relation for each
subcomponent on the right panel of Figure~\ref{fig:MassSize}. We
assign to each component a fraction of the total stellar mass
proportional to its flux.
Similarly, a power-law fit to the deV component yields $\log A$ and $\alpha$
of $0.50\pm 0.02$ and $0.74 \pm 0.09$.
While the slope remains unchanged within errors,
the mass-size relation for the deV component falls in between the relations at
$z=0.75$ and $z=0.25$ presented in \citet{2014ApJ...788...28V}.
The size measurements of \citet{2014ApJ...788...28V}
are corrected to be at the rest-frame wavelength $\lambda=5000$~\AA.
Although we do not correct our effective radii, they are measured in the
SDSS $r$-band images ($\lambda=6231$~\AA), so the $\approx 5\%$ correction cannot
explain the offset in the mass-size relation.
Thus, it is plausible that
the inner deV component in double component ETGs may be the relic of
the smaller, quiescent galaxies at $z\sim 0.5$ while the outer Exp
component assembled later, shifting the overall mass-size relation
towards the present-day relation.

\section{Discussion \& Summary} 
\label{sec:discussion}

In this study, we have performed a double component model fitting to a
sample of morphological elliptical galaxies in the nearby universe.
We have argued that the primary reason for preferring a double
component model to a single \Sersic\ model is the radially changing
shapes (ellipticities) of the isophotes, and selected ``double
component'' ETGs based on whether the double component model 
provides a better fit to the ellipticity profile.
We now examine whether and how
the results presented in the previous sections may be tied together in
a consistent scenario of the evolution of elliptical galaxies, in
association with the recent studies on ETGs.

First, we summarize the key findings of this paper as follows:
\begin{itemize}
  \item We find that a significant fraction (46 \%) of the nearby ETGs in our
  sample require multiple photometric components to describe their
  radial changes in isophotal shape. By fitting a two-component model
  to the surface brightness, we obtain that the characteristic sizes of the
  inner deV and outer Exp component are $\sim 3$~kpc and $\sim 20$~kpc
  respectively. The outer component tends to be more flattened than the
  inner, and this trend gets stronger with increasing
  luminosity.
  
  \item However, at the same time, both the shape and the flux
  fraction of the Exp component show significant spread at fixed stellar
  mass or luminosity. Interestingly, the inner component of the
  double-component galaxies follow the same fundamental plane relation
  derived from fitting a single component model for all ETGs, with
  similar scatter.

  \item Centrally concentrated (high \Sersic\ index), high $\sigma$ ETGs are 
    more likely to require a two-component fit to explain their 
    isophotal shape. The fraction of ETGs classified
    as double component shows a decreasing trend with the surface neighbor density
  at $\log\Sigma \gtrsim -0.5$~Mpc$^{-2}$ except for the highest $\sigma$ ETGs
  ($\sigma > 250$~km s$^{-1}$).

  \item The stellar populations of single and double component ETGs are
    different.  The double component ETGs generally tend to be
    younger and more metal-rich at $\sigma > 200$~km s$^{-1}$.

  \item The mass-size relation of the inner deV component in double component
    ETGs falls between the relation of quiescent galaxies at $z=0.25$ and $z=0.75$
    \citep{2014ApJ...788...28V}
    suggesting that the inner cores may have been in 
    place at z=0.75 and the outer components assembled 
    later.
\end{itemize}

The results of this study are in general agreement with the recent
multi-component fitting analysis of \citet{Huang:2013aa} as discussed
in section~\ref{sec:results}.  Because of the difference in image
quality, we do not resolve the inner component seen in their work with
median $R_e<1$~kpc, and the inner and middle component of their
decomposition would likely be jointly described by the deV in our
deVExp model. 

Both our work and \citet{Huang:2013aa} find
that the outer parts of nearby ETGs tend to be more elliptical than
the central component, and that this trend correlates with luminosity.
It is interesting to note that the even fainter and more extended
stellar halos of high concentration, massive galaxies seem to continue
this trend, showing higher ellipticity with increasing stellar mass
\citep{2014MNRAS.443.1433D}.  As opposed to \citet{Huang:2013ab}, who
found that multi-component fits are preferred in more than 75\% of the local
ETGs in their sample, we find a smaller fraction of galaxies require two 
components, and we uncover other interesting physical differences between 
the two samples. Our selection and the fraction of double
component ETGs in this work is more conservative in the sense that we
require a threshold change of \chisqe\ as well as a condition for the
second component to dominate the surface brightness at some region.

The inner deV component of our model with a typical $R_e\sim 3$~kpc moves
along the fundamental plane and shows a tight mass-size relation
(Figure~\ref{fig:fp}, \ref{fig:MassSize}). On the other hand,
the nature of the outer Exp components seem to be more stochastic.
They contribute a wide range in total galaxy light 
($\sim 0.3-0.8$) without any correlation with
the total stellar mass (Figure~\ref{fig:fdev}).
Even though their shapes tend to more elliptical with increasing luminosity,
the dispersion is still quite large (Figure~\ref{fig:ediff}).

The structures of ETGs are known to show connections with their
central stellar population
\citep[e.g.,][]{1998ApJ...508L..43F,2002MNRAS.330..547T,2009ApJ...698.1590G}.
The age, metallicity and abundances of ETGs strongly correlate with
$\sigma$ but not with size ($R_e$).  We also find that the
fraction of double component ETGs showing kpc-scale changes in isophote
shape increases with $\sigma$ but is independent of the overall size
typically measured by single component fitting.  Moreover these
studies suggest that the scatter about the fundamental plane may correlate
with age such that at fixed $\sigma$ and size, galaxies with higher mean
effective surface brightness are younger, more Fe-rich, and have lower
[Mg/Fe] than the counterparts with lower mean surface brightness
\citep{2010ApJ...717..803G,2010ApJ...721..278G}.
Our double component 
ETGs are analogous to the high surface brightness galaxies in the 
Graves et al.\ work, since the high $n$ values at fixed $\sigma$ that 
characterize these galaxies translates directly into higher surface 
brightness at fixed size. 
Likewise, the double component galaxies show younger 
ages and are more Fe-rich. What is new here is that we 
link high central surface brightness with a more prominent 
flattened outer component

We suggest as a possible scenario that the double component ETGs are
relics of relatively recent ($z\lesssim 1$) mass accretion onto the
outskirts of compact cores, whereas single component ETGs completed
this process earlier.  This is supported by the younger age and
under-dense environment for double component ETGs relative to the
single.
Older galaxies in denser environment may have experienced
a similar mass accretion
process at earlier times, but the signature of such events probed in this work
by radially changing ellipticity may be washed
out with time, leading us to classify them as single component ETGs.
Ongoing minor mergers in the double-component galaxies 
may drive the \Sersic\ index to a higher value by adding mass at 
large radius \citep{Hilz:2013aa}.  

From the perspective of the evolution of compact high-$z$ systems, the double
component ETGs may be those where the early core survives relatively intact.
\citet{2016MNRAS.456.1030W} studied 35 massive, compact galaxies at $z\approx2$
in the Illustris simulation \citep{2014MNRAS.444.1518V,2014MNRAS.445..175G}, a
suite of cosmological hydrodynamic simulations including baryonic physics and
sampling a large number of massive galaxies ($\Mstar=10^{9}-10^{12}~M_\odot$),
and suggest that the evolutionary paths of these systems are varied. The most
common outcome, occurring for roughly half of the sample, is that a compact,
massive galaxy survives as the core of a more massive galaxy at $z=0$, with the
outer envelope from merger and accretion.
This is roughly consistent with our findings.
On the other hand, in the simulation the compact cores at $z=2$ in denser
environment are more likely to gain stellar mass at larger radii
\citep[but also see][]{2015ApJ...815..104D}.
If all of the mass accumulation at large radius is detectable in our study
as changes in ellipticity profile, then the simulation results differ from
our finding that
the fraction of double component ETGs decrease in denser environment.
Thus far, there are no predictions on differences in the stellar population
for the different outcomes.

Recently, \citet{2016MNRAS.458.2371R} also looked at the in-situ mass fraction
of massive galaxies in the same simulation. They found that at fixed stellar
mass, galaxies hosted by halos with later formation times also have a higher
mass fraction in ex-situ stars than their early-formed counterparts. While in
their simulation they did not see a dependence on the mass-weighted stellar
age, it is still interesting to consider the possibility that we have found a
sub-population of galaxies with the highest ex-situ fraction, and that this
fraction appears to be linked with central stellar-population age.

Given that the double component ETGs have younger central stellar population
age, it is also possible that these galaxies joined the red sequence more
recently, and their outer components of higher ellipticities are from their
dissipational past rather than from size growth of the inner component (cores).
This would not necessarily support progenitor bias as the distribution of the
single-component model sizes of the two groups are similar.

The presence of extended light in galaxy outskirts has long been known to exist
ubiquitously for the brightest cluster galaxies (BCGs)
\citep[e.g.,][]{1976ApJ...209..693O,1988ApJ...328..475S,2005ApJ...618..195G}.
This ``cD envelope'' is thought to have assembled from the tidal interaction of
cluster galaxies
\citep{2004ApJ...607L..83M,2004MNRAS.355..159W,2005MNRAS.357..478S,2012ApJ...757...48M,2015MNRAS.451.2703C}
that forms intracluster light that traces the potential of the cluster. A
tantalizing similarity between the excess light around BCGs and massive ETGs is
that they are both largely aligned with the ``core'' component, and more
flattened \citep{1991AJ....101.1561P,2005ApJ...618..195G,2005MNRAS.358..949Z}.
Furthermore, in the case of BCGs, the orientation of the cluster galaxy
distribution is aligned with the BCGs \citep[e.g.,][]{1982A&A...107..338B}.
Taken together, the elongation of the cD envelope may be an imprint of the
disruption of galaxies falling into the cluster on primarily radial orbits
\citep{1996Natur.379..613M}.  This effect will be more pronounced at large
radial distances as the dynamical time is longer.  Although on a different
scale, we suspect a similar process might explain the elliptical outskirts that
we measure here around individual galaxies.

In this work, we modelled the surface brightness of nearby ETGs with a
multi-component model to argue that the large-scale radial changes in
ellipticity of many ETGs require a multi-component model to fully capture its
morphology.  We showed that this property may correlate with the properties of
the central stellar population, further hinting at the physical significance of
such decomposition.  In recent years, there have been significant efforts to
connect the kinematics of the central region of ETGs to their cosmological
evolution \citep[e.g.,][]{2014MNRAS.444.3357N}.  From the kinematics within
$\lesssim 1 R_e$, ETGs are commonly classified as either slow or fast rotators
\citep{2007MNRAS.379..401E,2011MNRAS.414..888E}.  The connection between such
classification, and our selection of double component ETGs is yet unclear.
Upcoming IFU surveys such as MANGA \citep{2015ApJ...798....7B} will be able to
study the connections between morphology, kinematics, and stellar population of
ETGs in greater detail, and ultimately how they are tied to the formation
history.

\acknowledgements

We thank Song Huang, Chung-Pei Ma and Kevin Bundy for useful discussions. We
also thank the referee for his/her detailed comments and suggestions that
significantly improved the paper. SO thanks Peter Erwin for his help with using
IMFIT. This research made use of Astropy, a community-developed core Python
package for Astronomy (Astropy Collaboration, 2013), and matplotlib
\citep{Hunter:2007}. JEG is partially supported by NSF grant AST-1411642.

\bibliographystyle{apj}
\bibliography{ref}

\end{document}